\documentclass[english,aps,prl,superscriptaddress,reprint]{revtex4-2}

\usepackage{graphicx}  
\usepackage{float}     
\usepackage{dcolumn}   
\usepackage{bm}        
\usepackage{amssymb}   
\usepackage{amsmath}
\usepackage{lipsum}    
\usepackage{xcolor}
\usepackage{soul}
\usepackage{float}
\usepackage{placeins}
\usepackage{physics}
\usepackage{mathrsfs}
\usepackage{braket}
\usepackage{subfigure}
\usepackage{bbold}
\usepackage{hyperref}
\hypersetup{
    colorlinks=true,     
    linkcolor=blue,      
    citecolor=blue,      
    urlcolor=blue,       
    }

\usepackage[utf8]{inputenc}
\usepackage{natbib}
\usepackage{graphicx}
\usepackage{subcaption}
\usepackage{booktabs}
\usepackage{blindtext}

\usepackage{color}

\definecolor{goodred}{RGB}{183,15,58}
\definecolor{goodblue}{RGB}{93,128,180}

\begin{document}

\title{
Spin Inertia as a Source of Topological Magnons: Chiral Edge States from Coupled Precession and Nutation}

\author{Subhadip Ghosh}
\affiliation{Department of Physics, Indian Institute of Technology (Indian School of Mines) Dhanbad, IN-826004, Dhanbad, India}

\author{Mikhail Cherkasskii}
\affiliation{Institute for Theoretical Solid State Physics, RWTH Aachen University, D-52074 Aachen, Germany}

\author{Ritwik Mondal}
\affiliation{Department of Physics, Indian Institute of Technology (Indian School of Mines) Dhanbad, IN-826004, Dhanbad, India}

\author{Alexander Mook}
\affiliation{Institute of Solid State Theory, University of M\"{u}nster, D-48149 M\"{u}nster, Germany}

\author{Levente Rózsa}
\affiliation{Department of Theoretical Solid State Physics, Institute for Solid State Physics and Optics, HUN-REN Wigner Research Centre for Physics, H-1525 Budapest, Hungary}
\affiliation{Department of Theoretical Physics, Budapest University of Technology and Economics, M\H{u}egyetem rkp. 3, H-1111 Budapest, Hungary}

\begin{abstract}
Spin inertia has been demonstrated to give rise to high-frequency nutational excitations beyond the conventional low-frequency precessional modes. Here, we demonstrate that the hybridization between precessional and nutational magnons may give rise to topological phenomena in the spin-wave spectrum. This hybridization requires the presence of interactions breaking angular-momentum conservation, such as the pseudodipolar interaction. We show on the example of a honeycomb ferromagnet how topological gaps open between the precessional and nutational bands that host chiral edge states in slab geometries. Our work establishes a theoretical foundation for exploring inertial spin dynamics as a new route to engineer topological phases in magnetic materials.
    
\end{abstract}
\maketitle

The rapid development in the generation of ultrashort laser pulses over the last decades has %
stimulated studies of magnetization dynamics on femto- to picosecond time scales~\cite{SIEGMANN1995L8,Beaurepaire1996,Kampfrath2023}. It has been proposed theoretically that the conventional Landau–Lifshitz–Gilbert (LLG) framework, which accounts for precession and damping, becomes incomplete at these time scales, and has to be extended by an inertial term~\cite{Ciornei2011,Suhl1998,Mondal2017Nutation,Olive2015,MONDAL2023170830}. This spin inertia induces a separation of the directions of the magnetic moment and the angular momentum, causing the former to perform a nutational motion around the latter. In ferromagnetic resonance spectra, spin inertia manifests as a secondary resonance peak appearing typically at terahertz frequencies~\cite{Olive2012}, which has been recently identified experimentally~\cite{neeraj2021inertial,unikandanunni2021inertial,De2025PRB}. Beyond modifying the linear response, inertial effects also influence spin transport~\cite{Mondal2021PRBSpinCurrent}, magnetization switching~\cite{Winter2022}, auto-oscillations~\cite{Rodriguez2024PRL}, and the spin–wave dispersion~\cite{Kikuchi2015,Mondal2022PRB,CherkasskiiPRB2024,Cherkasskii2021}. In the latter case, spin inertia results in the doubling of the excitation bands 
by giving rise to high-frequency nutational spin waves.

The topology of the band structure of spin waves or magnons has also attracted considerable attention recently~\cite{mcclarty2022}. Topological magnon insulators with chiral edge modes have been proposed theoretically in multi-sublattice ferromagnets~\cite{Zhang2013,Mook2014MHE,Mook2014ES,Chisnell2015}, magnon crystals~\cite{Shindou2013} and skyrmion lattices~\cite{RoldnMolina2016,Diaz2019,Weber2022}, and magnonic Weyl nodes~\cite{Li2016,Mook2016} have also been identified. The topology of the band structure also contributes to transport phenomena such as the magnon thermal Hall effect~\cite{Katsura2010,Matsumoto2011PRB,Matsumoto2011PRL,Onose2010}. A prototypical example of a topological magnon insulator is a ferromagnet on a honeycomb lattice, realized in, %
e.g., chromium trihalides and Cr(Si/Ge)Te$_{3}$ compounds~\cite{Chen2018,Chen2021,Zhu2021}. Two different microscopic mechanisms have been put forward that could explain the topologically non-trivial nature of the band structure in this geometry: the Dzyaloshinsky--Moriya interaction~\cite{DZYALOSHINSKY1958,Moriya1960,Chen2018,Chen2021} which is antisymmetric in the lattice sites, and various forms of the symmetric two-site exchange anisotropy including dipolar, pseudodipolar, and Kitaev terms~\cite{Wang2017,McClarty2018,Lee2020}. It has been suggested that the Dzyaloshinsky--Moriya interaction may be more prominent in the opening of a gap in the band structure~\cite{Jaeschke-Ubiergo2021}, since the gap size scales linearly with its strength but quadratically with the pseudodipolar term. Due to their different symmetry properties, it was also proposed that the two contributions may be distinguished by measuring the magnon spectrum for different orientations of the magnetization~\cite{Brehm2024}.

Here, we present how 
spin inertia influences the topology of the magnon band structure. As an example, we demonstrate that the precessional and nutational magnon bands may hybridize in a honeycomb ferromagnet, leading to an additional gap opening in the spectrum. This hybridization is enabled by the pseudodipolar interaction but not by the Dzyaloshinsky--Moriya interaction, enabling to distinguish between these two contributions, and the gap size is shown to scale linearly with the interaction strength. We confirm the topological nature of this gap by demonstrating the emergence of chiral edge modes in slab geometries, and connect this to the Chern numbers of the bands.

 The dynamics of the spins is described by the inertial Landau-Lifshitz-Gilbert (iLLG) equation~\cite{Ciornei2011}, 
\begin{align}
\dot{\bm{S}}_i=\bm{S}_i\times\left[-\gamma \bm{B}^{\rm eff}_i+\alpha\dot{\bm{S}}_i+\eta\ddot{\bm{S}}_i\right]\,.
\label{Eq2}
\end{align}
Here, $\bm{S}_i$ denotes a unit vector at site $i$ corresponding to the direction of the magnetic moment, $\gamma$ is the gyromagnetic ratio, $\alpha$ represents the Gilbert damping constant, and $\eta$ is the inertial relaxation time. The effective magnetic field is expressed as $\bm{B}^{\rm eff}_i = -\frac{1}{M_i}\frac{\partial \mathcal{H}}{\partial \bm{S}_i}$, where $\mathcal{H}$ is the Hamiltonian of the system and $M_{i}$ is the magnetic moment. The angular momentum at site $i$ may be defined as $\bm{L}_{i}=\frac{M_{i}}{\gamma}\left(\bm{S}_{i}-\eta\bm{S}_{i}\times\dot{\bm{S}}_{i}\right)$~\cite{MONDAL2023170830}, see the Supplemental Material (SM)~\cite{supp}. In the absence of an external magnetic field and spin--orbit coupling, the system is invariant under global rotations, which leads to the conservation of the total angular momentum $\sum_{i}\bm{L}_{i}$. 

To calculate the magnon excitations within linear spin-wave theory, first an equilibrium spin configuration is sought which satisfies $\bm{S}^{(0)}_i\times\bm{B}^{\rm eff}_i=\bm{0}$ at all lattice sites. Expanding Eq.~\eqref{Eq2} in small spin deviations $\bm{S}_{i}^{\bot}=\left[S_{i}^{-},S_{i}^{+}\right]=\left[\tilde{S}_{i}^{x}-\textrm{i}\tilde{S}_{i}^{y},\tilde{S}_{i}^{x}+\textrm{i}\tilde{S}_{i}^{y}\right]$, where in equilibrium $\tilde{S}_{i}^{z}=1$ at all sites in the local frames of reference, yields (cf. Ref.~\cite{CherkasskiiPRB2024})
\begin{align}\label{Eq5}
        \omega\begin{bmatrix}
        \bm{S}^{\bot}\\
        \bm{V}^{\bot}
    \end{bmatrix}=
    \begin{bmatrix}
        \mathbb{0}&\mathbb{1}\\
        \frac{\gamma}{\eta}\bm{M}^{-1}\mathcal{H}_{\textrm{SW}} &  \frac{\bm{\sigma}^z}{\eta}+\textrm{i}\frac{\alpha}{\eta}\mathbb{1}
    \end{bmatrix}
    \begin{bmatrix}
        \bm{S}^{\bot}\\
        \bm{V}^{\bot}
    \end{bmatrix}=\mathcal{D}\begin{bmatrix}
        \bm{S}^{\bot}\\
        \bm{V}^{\bot}
    \end{bmatrix}\,,
\end{align}
where $\bm{V}^{\bot}=-\textrm{i}\dot{\bm{S}}^{\bot}$ is the spin velocity, $\mathcal{H}_{\textrm{SW}}$ is the spin-wave Hamiltonian, $\mathbb{1}$ is the unit matrix, $\bm{M}$ is a diagonal matrix containing the magnetic moments, and $\bm{\sigma}^z$ acts as a Pauli matrix on the two components of $\bm{S}^{\bot}$. We substituted the harmonic time dependence $\bm{S}^{\bot}\propto\textrm{e}^{\textrm{i}\omega t}$, and replaced the time derivatives with multiplication by $\textrm{i}\omega$. The spin waves correspond
to the eigenmodes of the matrix $\mathcal{D}$ in Eq.~\eqref{Eq5}. The magnon frequencies can also be determined by solving the secular equation
\begin{align}\label{Eq6}
{\textrm{det}}\left(-\eta\omega^2\mathbb{1}+\omega\bm{\sigma}^z+\gamma\bm{M}^{-1}\mathcal{H}_{\rm SW}\right) = 0\,.
\end{align}
If the calculations are performed around a stable equilibrium state, $\mathcal{H}_{\textrm{SW}}$ is a positive definite operator, and the eigenvalue problem is quasi-Hermitian~\cite{Scholtz1992}. This means that all eigenvalues are real, but the left $\left<\bm{l}_{q}\right|$ and right $\left|\bm{r}_{q}\right>$ eigenvectors belonging to the same eigenvalue are not adjoints of each other, but form a biorthogonal system instead, $\left<\bm{l}_{q}|\bm{r}_{q'}\right>=\delta_{qq'}$. Furthermore, the eigenvalues are found in $\pm\omega_{q}$ pairs due to the particle-hole symmetry%
, which is represented as $\mathcal{C}=\textrm{diag}\left(\bm{\sigma}^{x},-\bm{\sigma}^{x}\right)\mathcal{K}$ in this system, where $\mathcal{K}$ denotes complex conjugation. All eigenvectors have the form $\left|\bm{r}_{q}\right>=\left[\bm{v}_{q},\omega_{q}\bm{v}_{q}\right]$, since the latter components are connected to the spin velocity.

\begin{figure}[H]
    \centering
\includegraphics[width=\columnwidth]{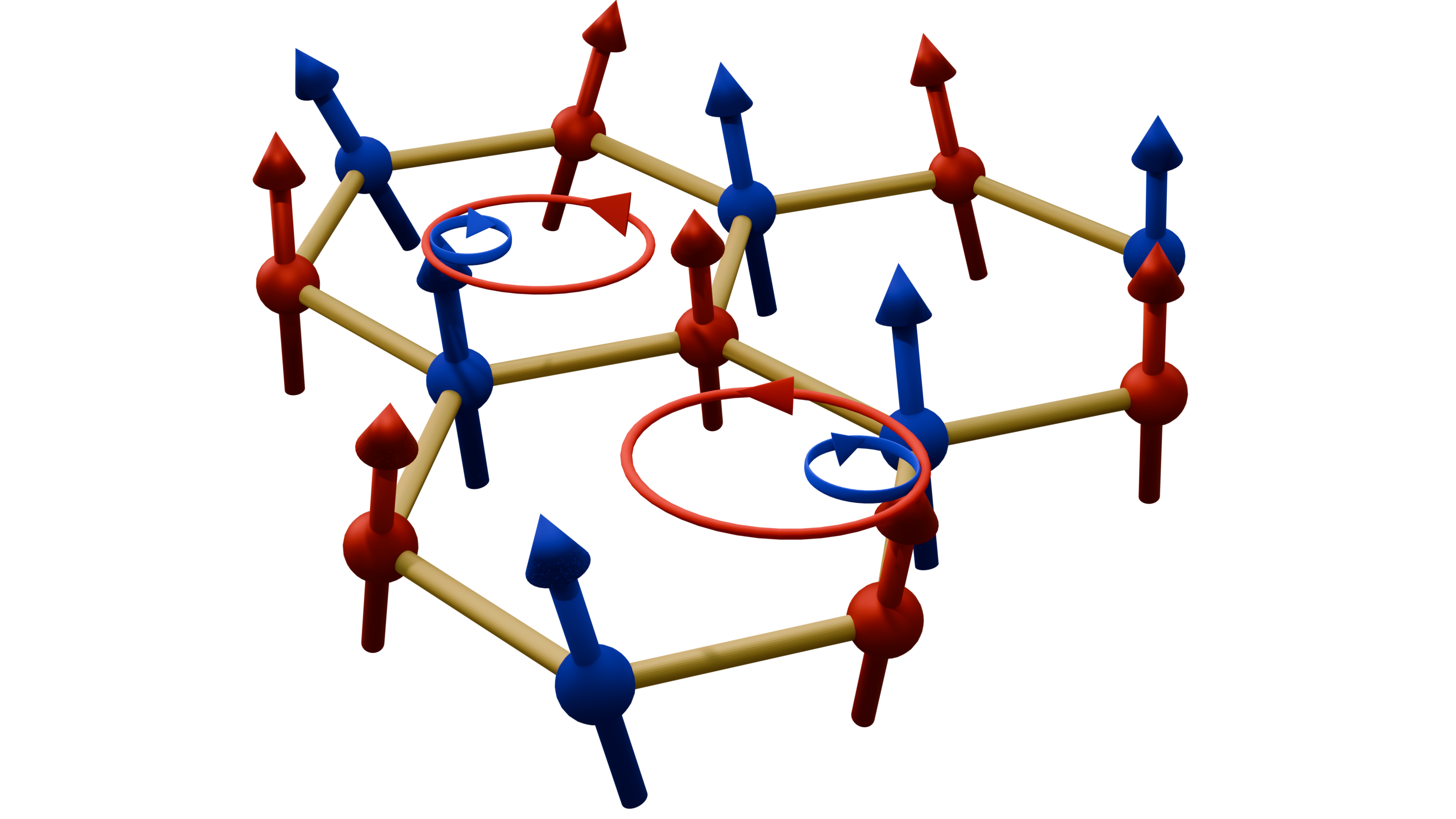}
\caption{Sketch of the ferromagnet on the honeycomb lattice, with red and blue arrows showing the two sublattices. Red and blue trajectories denote counterclockwise precessional and clockwise nutational modes, respectively. 
}
\label{Fig1}
\end{figure}

In the following, we consider a ferromagnetic system magnetized along the global $z$ direction, and describe the undamped modes for $\alpha=0$. The rotational symmetry around the $z$ axis is preserved in the absence of spin--orbit coupling, and the $z$ component of the total angular momentum, $\sum_{i}L^{z}_{i}$, remains a conserved quantity. On the level of spin-wave theory, this is reflected by the absence of pairing terms in the Hamiltonian $\mathcal{H}_{\textrm{SW}}$, which would couple the first and second components of $\bm{S}^{\bot}$ together. Considering only the $S_{i}^{+}$ sector, two types of excitations may be distinguished in inertial dynamics~\cite{Mondal_Ritwik_2021}: precessional excitations performing a counterclockwise rotation around the direction of the total angular momentum, which have $\omega>0$; and nutational excitations performing a clockwise rotation with $\omega<0$. These are illustrated in Fig.~\ref{Fig1}. Due to the different rotational senses, the angular momenta of the excited states also differ: $\sum_{i}L^{z}_{i}>\sum_{i}\frac{M_{i}}{\gamma}S^{z}_{i}$ for precessional and $\sum_{i}L^{z}_{i}<\sum_{i}\frac{M_{i}}{\gamma}S^{z}_{i}$ for nutational motion. To observe topological edge modes connecting the precessional and nutational magnon bands, this conservation of the total angular momentum has to be broken.

As an example, we consider a ferromagnet on the honeycomb lattice. The Hamiltonian is given by
\begin{align}\label{Eq1}
    \mathcal{H}  = \sum_{\langle i, j\rangle} \left[\bm{S}^A_i \bm{J}_{ij} \bm{S}^B_j +F \left(\bm{S}_i^A\cdot\hat{\bm{r}}_{ij}\right)\left(\bm{S}_j^B\cdot \hat{\bm{r}}_{ij}\right)\right]\,,
\end{align}
where $A$ and $B$ indicate the two sublattices denoted by red and blue spins in Fig.~\ref{Fig1}, respectively. The exchange interaction tensor $\bm{J}_{ij}$ between nearest neighbors is diagonal here, with $J^{zz}<J^{xx}=J^{yy}<0$ preferring magnetization along $\hat{\bm{z}}$. The pseudodipolar interaction $F$ connects the spin directions to the relative positions of the sites via the unit vector $\hat{\bm{r}}_{ij}$ connecting sites $i$ and $j$, which is either $\hat{{\bm \delta}}_1= \left(-1,0\right)$, $\hat{\bm{\delta}}_2=\left(\frac{1}{2},-\frac{\sqrt{3}}{2}\right)$, or $\hat{\bm{\delta}}_3=\left(\frac{1}{2},\frac{\sqrt{3}}{2}\right)$. The parameter values used in the simulations are summarized in Table~\ref{tab:1}. The exchange interaction $J^{xx}=-3$~meV and the magnetic moments $M_{A}=M_{B}=3\mu_{\textrm{B}}$ are comparable to experimentally determined values for honeycomb van der Waals materials~\cite{Chen2018,Soriano2020,Chen2021,Zhu2021}. The inertial relaxation time 
was set to $\eta=330$~fs, which is similar to the values determined experimentally in transition metals~\cite{neeraj2021inertial,MONDAL2023170830}. 

\begin{table}[H]
    \caption{Dimensionless model parameters used throughout the text.}
   \centering
    \begin{tabular}{ c | c | c | c | c |c }
    \hline
    \hline 
    
         $J^{xx}$ (meV) & $J^{zz}$ (meV)  & $F$ (meV)&  $\eta$ (fs)& $M_{A}=M_{B}$ $(\mu_{\textrm{B}})$  \\
         
        \hline 
         $-3.0$&  $-3.3$ &  $1.5$ &$330$  &$3$  \\
         \hline\hline
    \end{tabular}
    \label{tab:1}
\end{table}

The spin-wave Hamiltonian, after spatial Fourier transformation $\bm{S}^{A/B,\bot}\left(\bm{k}\right)=N^{-1/2}\sum_{i/j}\textrm{e}^{-\textrm{i}\bm{k}\bm{R}_{i/j}}\bm{S}_{i/j}^{A/B,\bot}$, may be expressed as
 \begin{align}
    \mathcal{H}_{\textrm{SW},\bm{k}}=& \begin{pmatrix}
         \Omega & \mathcal{H}_{1}& 0 &\tilde{F}_{1}-{\rm i}\tilde{F}_{2}\\[10pt]
         \mathcal{H}_{1}^{*} & \Omega & \tilde{F}_{1}^{*}-{\rm i}\tilde{F}^*_{2} & 0\\[10pt]
         0 &\tilde{F}_{1}+{\rm i}\tilde{F}_{2} & \Omega & \mathcal{H}_{1}\\[10pt]
         \tilde{F}_{1}^{*}+{\rm i}\tilde{F}^*_{2} & 0 &\mathcal{H}_{1}^{*} & \Omega
      \end{pmatrix}\,,
\end{align}
where the components are written in the order $\left[S_{-}^{A},S_{-}^{B},S_{+}^{A},S_{+}^{B}\right]$. The matrix elements are given by
\begin{align}
\Omega=&-3J^{zz},\\
\mathcal{H}_{1}\left(\bm{k}\right)=&\left(J^{xx}+\frac{F}{2}\right)\left(\textrm{e}^{\textrm{i}\bm{k}\bm{\delta}_{1}}+\textrm{e}^{\textrm{i}\bm{k}\bm{\delta}_{2}}+\textrm{e}^{\textrm{i}\bm{k}\bm{\delta}_{3}}\right),\\
\tilde{F}_{1}\left(\bm{k}\right)=&\frac{F}{4}\left(2\textrm{e}^{\textrm{i}\bm{k}\bm{\delta}_{1}}-\textrm{e}^{\textrm{i}\bm{k}\bm{\delta}_{2}}-\textrm{e}^{\textrm{i}\bm{k}\bm{\delta}_{3}}\right),\\
\tilde{F}_{2}\left(\bm{k}\right)=&\frac{\sqrt{3}F}{4}\left(\textrm{e}^{\textrm{i}\bm{k}\bm{\delta}_{3}}-\textrm{e}^{\textrm{i}\bm{k}\bm{\delta}_{2}}\right).
\end{align}
 
\begin{figure}[H]
    \centering
\includegraphics[width=\columnwidth]{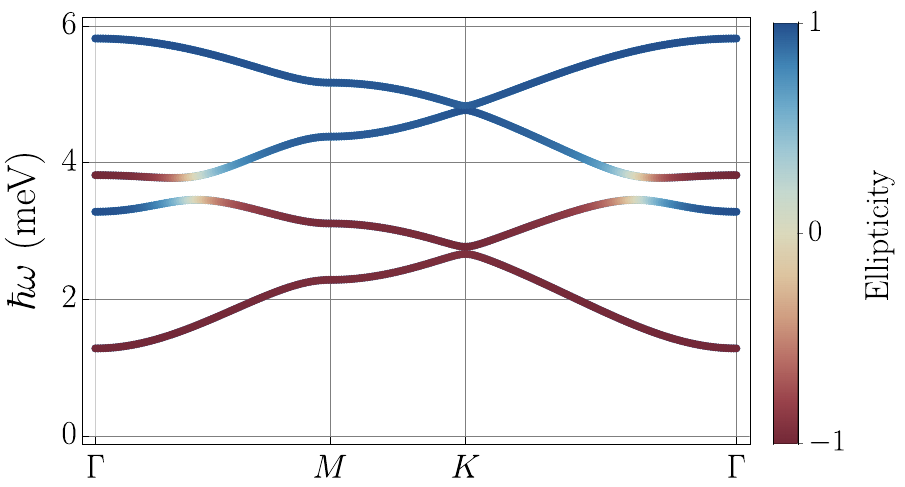}
\caption {Magnon band structure in momentum space along high-symmetry points. The states are colored according to their ellipticity defined in Eq.~\eqref{Eq6}. 
}
\label{Fig2}
\end{figure}

The calculated magnon spectrum is shown in Fig.~\ref{Fig2} along the high-symmetry lines of the Brillouin zone. The exchange interaction forms two precessional bands in this two-sublattice structure, which touch in the $K$ and $K'$ points forming Dirac cones. Spin inertia leads to additional nutational modes, which have the same dispersion apart from being shifted up in frequency by $\eta^{-1}$~\cite{Mondal2022PRB}; see the SM~\cite{supp}. For sufficiently high values of $\eta\gamma M^{-1}\left|J^{xx}\right|$, first the top of the higher precessional band touches the bottom of the lower nutational band in the $\Gamma$ point, then a nodal line is formed between the bands along a contour encompassing the $\Gamma$ point. The pseudodipolar interaction $F$ breaks the continuous rotational symmetry around $\hat{\bm{z}}$, thereby causing a hybridization between the precessional and nutational modes and opening gaps in the spectrum. In the Dirac points, the gap is opened for any finite value of $F$, but its size scales with $\left|F\right|^{2}$, making it difficult to identify in Fig.~\ref{Fig2}; see the SM~\cite{supp}. Along the nodal line, the gap size scales linearly with $\left|F\right|$, but this effect can only be observed if the inertial relaxation time is sufficiently high such that the bands overlap. The hybridization between the nutational and precessional bands may be captured by calculating the ellipticity of mode $n$ at wave vector $\bm{k}$,
\begin{align}
e_{n,\bm{k}}=2\left<\bm{l}_{n,\bm{k}}\right|P_{-}\left|\bm{r}_{n,\bm{k}}\right>-1,\label{Eq6}
\end{align}
where $P_{-}=\textrm{diag}\left(1,1,0,0,1,1,0,0\right)$ is the projection onto the clockwise rotating $S_{-}-V_{-}$ subspace. The ellipticity takes the value of $1$ for left circularly polarized, typically nutational, modes, and $-1$ for right circularly polarized, typically precessional, modes, while it vanishes for linearly polarized modes forming where the degeneracy along the nodal line is lifted. Alternatively, $e_{n,\bm{k}}$ may be thought of as the angular momentum carried by the mode, as discussed above. The avoided crossing may also be detected by studying the polarizations of the modes only in the high-symmetry points: in the $\Gamma$ point where the right-handed and left-handed modes are eigenstates, the sign of the ellipticity is alternating with increasing energy, while in the $M$ and $K$ points two predominantly right-handed modes are followed by two mainly left-handed modes; see the Supplemental Movies~\cite{supp}.

\begin{figure}[H]
    \centering
    \includegraphics[width=\columnwidth]{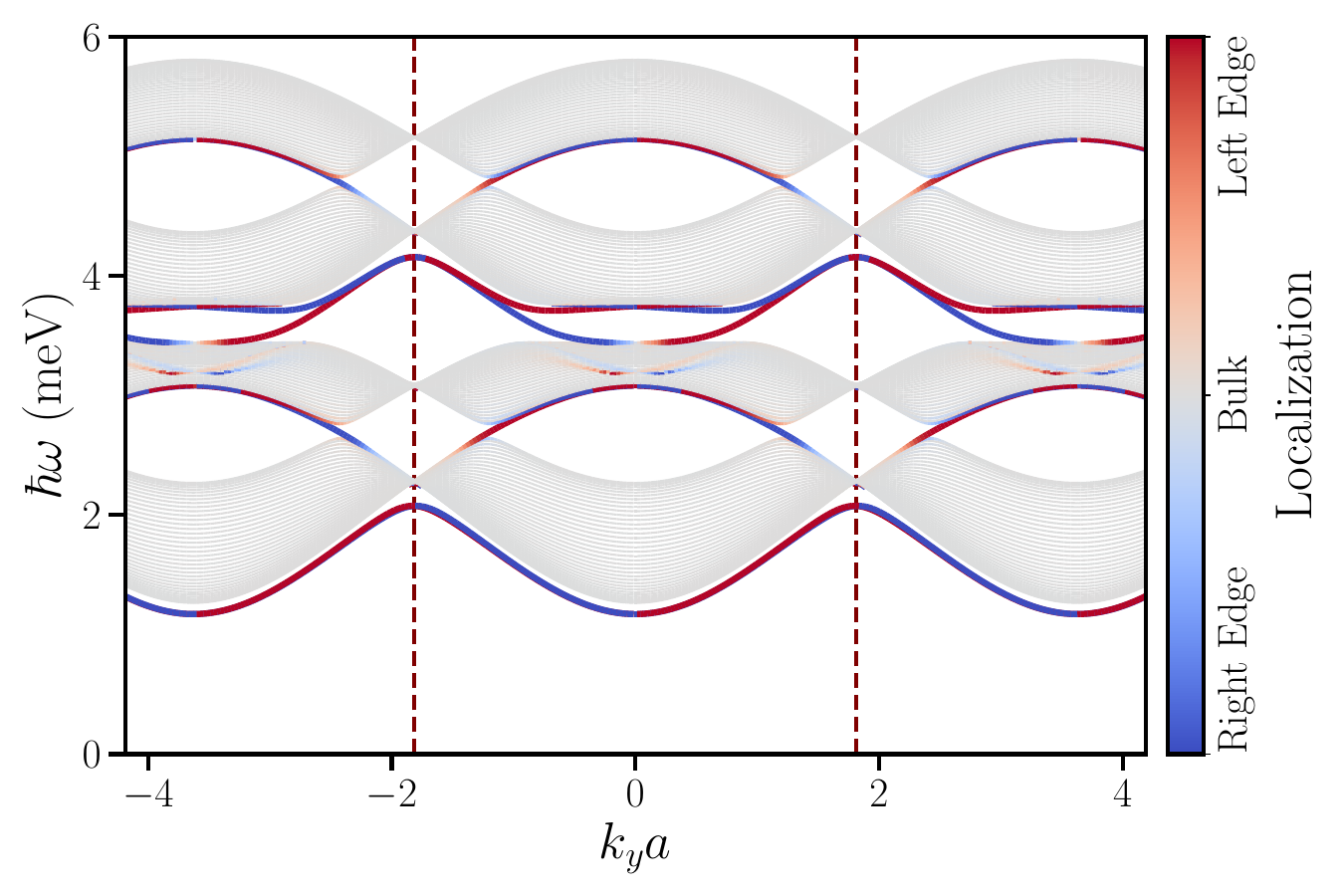}
    \caption{Magnon band structure in a slab geometry. The system is $40$ unit cells wide along the $x$ direction with open boundary conditions, while it is infinite along the $y$ direction with periodic boundary conditions, meaning that the eigenstates can be indexed by the wave vector component $k_{y}$. $a$ denotes the distance between nearest-neighbor atoms in the lattice, i.e., the lattice constant is $\sqrt{3}a$. Dotted lines show the boundaries of the one-dimensional Brillouin zone. Coloring shows the localization defined in Eq.~\eqref{Eq7}. 
    }
    \label{Fig3}
\end{figure}

The gaps opened in the $K$ and $K'$ points have been demonstrated to be topologically non-trivial in this model~\cite{Wang2017}, and this may also be the case for the gap between the precessional and nutational bands. A typical signature of topologically non-trivial gaps is the presence of chiral edge modes at the boundaries~\cite{Mook2014ES}. To confirm this bulk-edge correspondence, we recalculated the magnon spectrum in a slab geometry, as shown in Fig.~\ref{Fig3}. The edge modes are identified by calculating the localization
\begin{align}
\ell_{n,\bm{k}}=\left<\bm{l}_{n,\bm{k}}\right|P_{\textrm{Left}}\left|\bm{r}_{n,\bm{k}}\right>-\left<\bm{l}_{n,\bm{k}}\right|P_{\textrm{Right}}\left|\bm{r}_{n,\bm{k}}\right>,\label{Eq7}
\end{align}
where $P_{\textrm{Left}}$ and $P_{\textrm{Right}}$ are projections onto the sites on the left and right half of the slab, respectively. Close to the Dirac cones where the first and second, respectively the third and fourth bands almost touch, there is a single edge mode crossing the gap. These modes are localized at opposite edges of the system, and propagate in opposite directions based on their group velocity. Chiral edge modes may also be observed between the second and third bands where the avoided crossing between the precessional and nutational bands is found in Fig.~\ref{Fig2}. While there is a single mode localized on each edge in this gap as well, note that the group velocity changes sign along these modes, with the edge modes having a frequency maximum at the edge of the one-dimensional Brillouin zone. This indicates that the gap between any adjacent pairs of bands is topologically non-trivial. Note that a pair of edge modes is also observable below the lowest magnon band. This is due to the reduced number of neighbors at the edge, which destabilizes the ferromagnetic alignment along the $z$ direction for higher values of the pseudodipolar interaction than what is considered here. However, these edge modes are of topologically trivial origin, and do not merge with the bulk bands as the wave vector varies.

\begin{figure}[H]
    \includegraphics[width=\columnwidth]{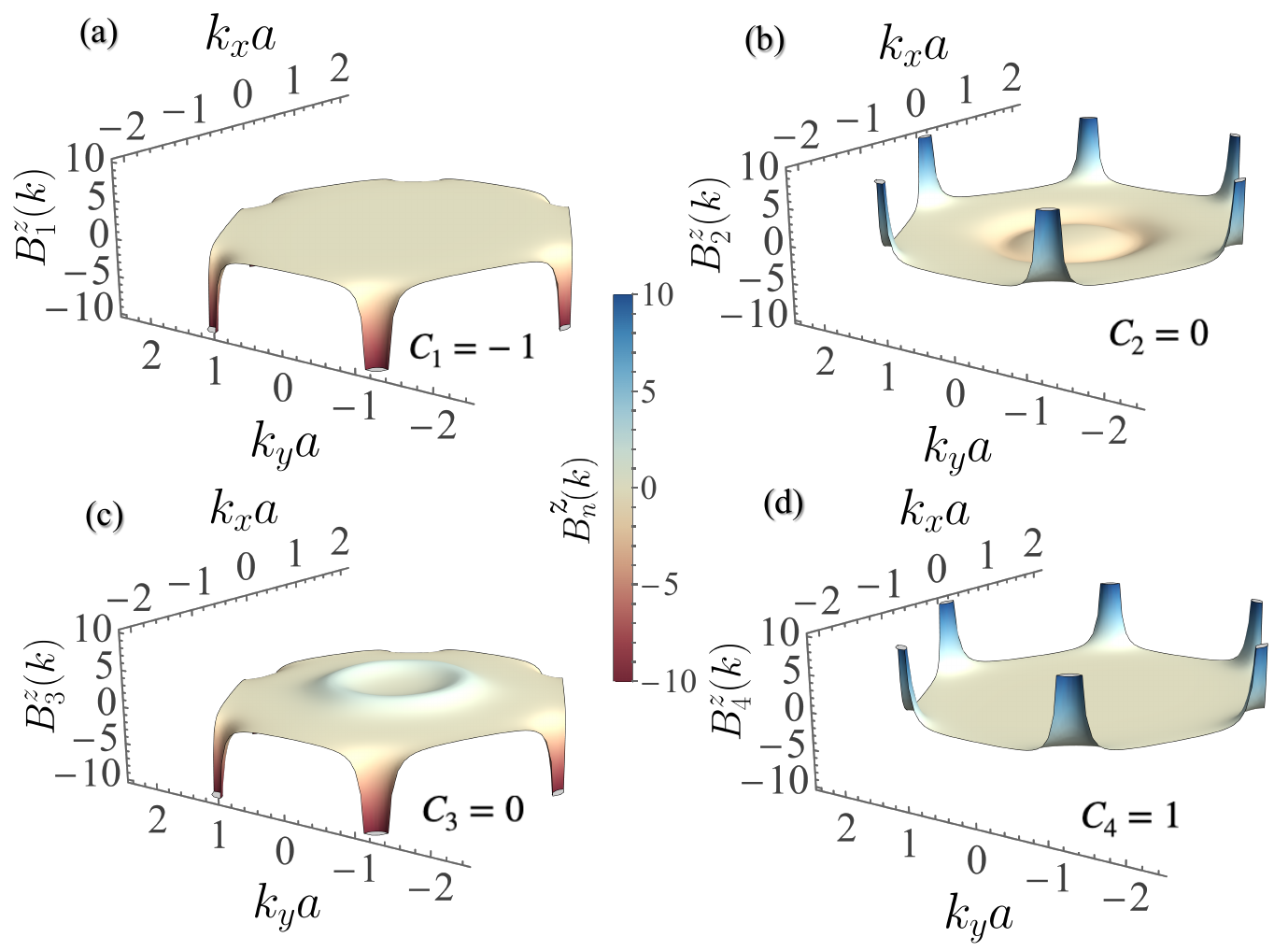}
    \caption{Berry curvature $B^{z}_{n}(\bm{k})$ of the bands calculated from Eq.~\eqref{Eq8}. (a)-(d) show the four magnon bands in increasing order of frequency. The Chern numbers $C_{n}$ for each band are also given. 
    }
    \label{Fig4}
\end{figure}

To confirm the topologically non-trivial origin of the edge states, we calculate the topological invariants of the bands in the infinite system, corresponding to the Chern number in the considered case. The Chern number is the integral of the $z$ component of the Berry curvature $C_{n}=\left(2\pi\right)^{-1}\int B^{z}_{n}\left(\bm{k}\right)\textrm{d}^{2}\bm{k}$, where the latter may be expressed with the left and right eigenvectors as (see the SM~\cite{supp} for details)
\begin{align}\label{Eq8}
    B^{\alpha}_{n}\left(\bm{k}\right) =
    {\rm i}\epsilon^{\alpha\beta\gamma}\sum_{m\neq n}\frac{\braket{\bm{l}_{n,\bm{k}}|\partial_{k^{\beta}}\mathcal{D}_{\bm{k}}|\bm{r}_{m,\bm{k}}}\braket{\bm{l}_{m,\bm{k}}|\partial_{k^{\gamma}}\mathcal{D}_{\bm{k}}|\bm{r}_{n,\bm{k}}}}{\left(\omega_{n,\bm{k}}-\omega_{m,\bm{k}}\right)^2}.
\end{align}
Here, $\mathcal{D}_{\bm{k}}$ is the matrix on the right-hand side of Eq.~\eqref{Eq5} 
in Fourier space, and $\alpha,\beta,\gamma$ denote Cartesian indices. The distribution of $B^{z}_{n}$ for the four bands is shown in Fig.~\ref{Fig4}. The sum of the Chern numbers $\left|\sum_{i=1}^{n}C_{i}\right|$ predicts the number of pair of chiral edge modes between bands $n$ and $n+1$, which equals $1$ between any two bands, in agreement with Fig.~\ref{Fig3}. The Chern number $C_{1}=-1$ of the lowest band in Fig.~\ref{Fig4}(a) stems from the peaks in the Berry curvature close to the $K$ and $K'$ points where the degeneracy is lifted. These peaks are also observable in the second band in Fig.~\ref{Fig4}(b) with opposite sign, but they are cancelled out by the circular depression around the contour where the precessional and nutational bands overlap in the absence of spin--orbit coupling. The Berry curvatures in the third and fourth bands are qualitatively similar to the second and first bands, respectively, although with reversed signs.

In summary, we demonstrated that the hybridization between nutational and precessional spin waves in inertial spin dynamics may modify the topological properties of the bands. We illustrated this on the example of a ho\-neycomb ferromagnet, which realizes a magnonic Chern insulator with an avoided crossing between the precessional and nutational bands in the bulk system, which is filled by chiral edge modes in a slab geometry. 
 
Note that the width of the magnon bands is lower in the calculations than what was experimentally determined in van der Waals magnets~\cite{Chen2018,Soriano2020,Chen2021,Zhu2021}; this is in part caused by the compression of the bands by the inertial term~\cite{Mondal2022PRB}, as shown in the SM~\cite{supp}. This means that the value of the exchange interaction may be higher than what was determined from the non-inertial model, and since the topological transition is governed by the parameter $\gamma M^{-1}\left|J^{xx}\right|\eta$, it may be observed for lower inertial relaxation times than the value used in this work. The inertial relaxation time may be further enhanced in the presence of heavy elements with high spin--orbit coupling in these materials. Therefore, the proposed effects may be experimentally accessible by measuring the magnon spectrum in the whole Brillouin zone up to higher energies where the nutational bands are expected to emerge.

The nutational and precessional bands may be distinguished based on their different angular momentum, or opposite rotational sense. This means that methods sensitive to the rotational sense of the excitations may be particularly suitable for this purpose, such as spin-polarized electron-energy loss spectroscopy. Optical methods may also be applied in the vicinity of the $\Gamma$ point where the reversed rotational sense of the higher modes already provides an indication for the topologically non-trivial nature of the bands. 
Detecting the inertial dynamics may also shed further light on the role of the competing mechanisms of the Dzyaloshinsky--Moriya interaction and the pseudodipolar interaction in the gap opening in these systems~\cite{Chen2021,Brehm2024}: while the pseudodipolar interaction may open a gap between the precessional and nutational bands, the out-of-plane component of the Dzyaloshinsky--Moriya interaction has no such effect since it preserves the angular momenta of the excitations.

These calculations may be generalized to other systems, for example single-sublattice ferromagnets, where topologically non-trivial bands may only be formed if inertial effects are taken into account; see the SM~\cite{supp} for an example. The generation of bulk and topological edge modes in ferromagnets at THz frequencies may also open further perspectives in controlling the angular-momentum transfer between magnons and phonons or electrons.

\begin{acknowledgments}
    A.M. acknowledges funding from the Deutsche Forschungsgemeinschaft (DFG, German Research Foundation) through TRR 173-268565370 (Project B13), and Project No.~504261060 (Emmy Noether Programme), and from the Dynamics and Topology Center (TopDyn) funded by the State of Rhineland-Palatinate. R.M. and A.M. acknowledge the Indo-German Science Technology Centre (IGSTC) for the exchange visit via the PECFAR award No. IGSTC/PECFAR/Call 2024/IGSTC-02232/RM-AM/70/2024-25/74.
    M.C. acknowledges funding from the Deutsche Forschungsgemeinschaft (DFG, German Research Foundation) under the FOR 5844: ChiPS, Project-ID 541503763. L.R. gratefully acknowledges funding by the National Research, Development, and Innovation Office (NRDI) of Hungary under Project Nos. FK142601 and ADVANCED 149745, and by the Hungarian Academy of Sciences via a J\'{a}nos Bolyai Research Grant (Grant No. BO/00178/23/11).
\end{acknowledgments}

\clearpage
\onecolumngrid

\setcounter{equation}{0}
\setcounter{figure}{0}
\setcounter{table}{0}
\setcounter{section}{0}

\renewcommand{\theequation}{S\arabic{equation}}
\renewcommand{\thefigure}{S\arabic{figure}}
\renewcommand{\thetable}{S\arabic{table}}
\renewcommand{\thesection}{S\arabic{section}}

\setcounter{NAT@ctr}{0}
\renewcommand{\bibnumfmt}[1]{[S#1]}


\begin{center}
{\bfseries\large Supplementary Material to}\\[6pt]
{\large  \bfseries Spin Inertia as a Source of Topological Magnons: Chiral Edge States from Coupled Precession and Nutation}\\[6pt]
Subhadip Ghosh$^1$, Mikhail Cherkasskii$^2$, Ritwik Mondal$^1$, Alexander Mook$^3$, Levente Rózsa$^{4,5}$\\
\vspace{0.5cm}
{\it \small $^1$Department of Physics, Indian Institute of Technology (Indian School of Mines) Dhanbad, IN-826004, Dhanbad, India\\
$^2$Institute for Theoretical Solid State Physics, RWTH Aachen University, D-52074 Aachen, Germany\\
$^3$Institute of Solid State Theory, University of M\"{u}nster, D-48149 M\"{u}nster, Germany\\
$^4$Department of Theoretical Solid State Physics, Institute for Solid State Physics and Optics, HUN-REN Wigner Research Centre for Physics, H-1525 Budapest, Hungary\\
$^5$Department of Theoretical Physics, Budapest University of Technology and Economics, M\H{u}egyetem rkp. 3, H-1111 Budapest, Hungary}

\vspace{0.5cm}

Here, we present considerations about angular-momentum conservation, the derivation of linear spin-wave theory and the Berry curvature in the inertial regime, a discussion of the parameter dependence of the size of the topological gap and of the topological phase transition in the honeycomb lattice, and the formation of a magnonic Chern insulating state in a single-sublattice ferromagnet.
\end{center}


\twocolumngrid
\section{Angular-momentum conservation}

The Hamiltonian $\mathcal{H}$ is invariant under global rotations if $\mathcal{H}\left(\left\{\bm{S}_{i}\right\}\right)=\mathcal{H}\left(\left\{R\bm{S}_{i}\right\}\right)$, where $R\in\textrm{SO}(3)$ is an arbitrary rotation matrix. This implies that the Hamiltonian only depends on scalar products of spins on different sites, $\mathcal{H}\left(\left\{\bm{S}_{i}\right\}\right)=\mathcal{H}\left(\left\{\bm{S}_{i}\cdot\bm{S}_{j}\right\}\right)$, similarly how from translational invariance in classical mechanics it follows that the Hamiltonian may only depend on relative positions. Consider a single scalar-product term $\mathcal{H}_{1}=\bm{S}_{i}\cdot\bm{S}_{j}$. From this it follows that
\begin{align}
\bm{S}_{i}\times\frac{\partial \mathcal{H}_{1}}{\partial \bm{S}_{i}}+\bm{S}_{j}\times\frac{\partial \mathcal{H}_{1}}{\partial \bm{S}_{j}}=\bm{S}_{i}\times\bm{S}_{j}+\bm{S}_{j}\times\bm{S}_{i}=\bm{0},\label{EqS1}
\end{align}
implying
\begin{align}
\sum_{k}\bm{S}_{k}\times\frac{\partial \mathcal{H}_{1}}{\partial \bm{S}_{k}}=\bm{0}.\label{EqS2}
\end{align}
This relation is straightforward to generalize to any rotationally symmetric Hamiltonian $\mathcal{H}$ via the chain rule.

Equation~(1) in the main text may be rearranged as
\begin{align}
\textrm{d}_{t}\left(\bm{S}_i-\eta\bm{S}_i\times\dot{\bm{S}}_i\right)=\bm{S}_i\times\left[-\gamma \bm{B}^{\rm eff}_i+\alpha\dot{\bm{S}}_i\right]\,,
\label{EqS3}
\end{align}
since $\dot{\bm{S}}_i\times\dot{\bm{S}}_i=\bm{0}$. Setting $\alpha=0$ and multiplying the equation by $M_{i}/\gamma$, then summing over the sites yields
\begin{align}
\textrm{d}_{t}\sum_{i}\frac{M_{i}}{\gamma}\left(\bm{S}_i-\eta\bm{S}_i\times\dot{\bm{S}}_i\right)=\sum_{i}\bm{S}_i\times\frac{\partial \mathcal{H}}{\partial \bm{S}_{i}}.\label{EqS4}
\end{align}
For a rotationally invariant system, the right-hand side of the equation vanishes as shown in Eq.~\eqref{EqS2}. This leads to the conservation of the sum of the quantities $\bm{L}_{i}=\frac{M_{i}}{\gamma}\left(\bm{S}_{i}-\eta\bm{S}_{i}\times\dot{\bm{S}}_{i}\right)$, which is identified as the angular momentum, the conserved quantity traditionally associated with rotational invariance. For an alternative motivation of this expression as an angular momentum, see Ref.~\cite{Ciornei2011_supp}.

\section{Linear spin-wave theory in the inertial regime}

Equation~(2) in the main text may be derived by rewriting Eq.~(1) as a coupled set of first-order differential equations,
\begin{align}
\dot{\bm{S}}_i  =& \textrm{i}\bm{V}_i\,,\label{EqS5}\\
    \dot{\bm{V}}_i  =& -\frac{1}{\eta}\bm{S}_i\times\bm{V}_i+\textrm{i}\frac{\gamma}{\eta}\left[\bm{S}_i\left(\bm{S}_i\cdot\bm{B}^{\rm eff}_i\right)-\bm{B}^{\rm eff}_i\right]-\frac{\alpha}{\eta}\bm{V}_i\nonumber\\
    &-\textrm{i}\bm{S}_i\bm{V}^2_i,
    \label{EqS6}
\end{align}
by introducing $\bm{V}=-\textrm{i}\dot{\bm{S}}$ as an additional variable. The equation has the same form when expressed in the local coordinate system with the spin vectors $\tilde{\bm{S}}_{i}$. The Hamiltonian $\mathcal{H}$ is expanded around the equilibrium state $\tilde{S}_{i}^{z}=1$ up to second order in the small variables $\bm{S}^{\bot}$ as
\begin{align}
\mathcal{H}\approx E_{0}+\frac{1}{4}\left(\bm{S}^{\bot}\right)^{\dagger}\mathcal{H}_{\rm SW}\bm{S}^{\bot}\,.\label{EqS7}
\end{align}
Note that $\bm{S}^{\bot}$ has $\bm{S}^{-}$ and $\bm{S}^{+}$ components, and is also a vector over the lattice sites. The spin-wave Hamiltonian is characterized by the particle-hole constraint $\bm{\sigma}^{x}\mathcal{K}\mathcal{H}_{\rm SW}\mathcal{K}\bm{\sigma}^{x}=\mathcal{H}_{\rm SW}$, expressing that $\bm{S}^{-}$ and $\bm{S}^{+}$ are complex conjugates of each other as follows from their definition.

In Eqs.~\eqref{EqS5} and \eqref{EqS6}, we only keep terms up to linear order in the small variables $\bm{S}^{\bot}$. This implies that the velocities $\bm{V}^{\bot}$ are also treated as small variables, while we substitute $\tilde{S}_{i}^{z}=1$ and consequently $\tilde{V}_{i}^{z}=0$ for the third components. This simplifies the equation to
\begin{align}
        \begin{bmatrix}
        \dot{\bm{S}}^{\bot}\\
        \dot{\bm{V}}^{\bot}
    \end{bmatrix}=
    \textrm{i}\begin{bmatrix}
        \mathbb{0}&\mathbb{1}\\
        \frac{\gamma}{\eta}\bm{M}^{-1}\mathcal{H}_{\textrm{SW}} &  \frac{\bm{\sigma}^z}{\eta}+\textrm{i}\frac{\alpha}{\eta}\mathbb{1}
    \end{bmatrix}
    \begin{bmatrix}
        \bm{S}^{\bot}\\
        \bm{V}^{\bot}
    \end{bmatrix}=\textrm{i}\mathcal{D}\begin{bmatrix}
        \bm{S}^{\bot}\\
        \bm{V}^{\bot}
    \end{bmatrix}\,,\label{EqS8}
\end{align}
and replacing the time derivative with $\textrm{i}\omega$ yields the eigenvalue equation (2) in the main text. Note that the $-\textrm{i}$ prefactor in the definition of $\bm{V}$ was introduced such that there is no explicit $\textrm{i}$ factor in the $\mathcal{D}$ matrix for $\alpha=0$, and the first and second parts of the eigenvectors are connected by multiplication via the real frequencies $\omega$.

Introducing the notations
\begin{align}
\mathcal{B}=&\frac{\gamma}{\eta}\bm{M}^{-1},\label{EqS9}\\
\mathcal{A}=&\left[\mathcal{B}^{\frac{1}{2}}\mathcal{H}_{\textrm{SW}}\mathcal{B}^{\frac{1}{2}}\right]^{\frac{1}{2}},\label{EqS10}\\
\mathcal{F}=&\begin{bmatrix}\mathcal{A}\mathcal{B}^{-\frac{1}{2}}&\mathbb{0}\\\mathbb{0}&\mathcal{B}^{-\frac{1}{2}}\end{bmatrix},\label{EqS11}
\end{align}
one obtains the relation
\begin{align}
\mathcal{D}=\mathcal{F}^{-1}\begin{bmatrix}\mathbb{0}&\mathcal{A}\\\mathcal{A}& \frac{\bm{\sigma}^z}{\eta}\end{bmatrix}\mathcal{F}.\label{EqS12}
\end{align}
This means that $\mathcal{D}$ is connected to a Hermitian matrix by a basis transformation $\mathcal{F}$, i.e., it is quasi-Hermitian~\cite{Scholtz1992_supp}. Note that the matrix $\mathcal{B}$ is always strictly positive definite, and $\mathcal{A}$ also needs to be strictly positive definite to be invertible. While the spin-wave Hamiltonian $\mathcal{H}_{\textrm{SW}}$ is necessarily positive semidefinite when the expansion is carried out around a stable equilibrium state, it may be necessary to project out its subspace belonging to zero eigenvalue to perform this transformation. The quasi-Hermiticity implies that the frequencies $\omega$ are real valued, since they are not influenced by the basis transformation, while the particle-hole constraint of the spin-wave Hamiltonian enforces $\mathcal{C}\mathcal{D}\mathcal{C}^{-1}=-\mathcal{D}$ for $\mathcal{C}=\textrm{diag}\left(\bm{\sigma}^{x},-\bm{\sigma}^{x}\right)\mathcal{K}$, meaning that the non-zero frequencies are found in $\pm \omega$ pairs.

The left and right eigenvectors of $\mathcal{D}$ are connected via
\begin{align}
\left<\bm{l}_{q}\right|=\left|\bm{r}_{q}\right>^{\dagger}\mathcal{F}^{2},\label{EqS13}
\end{align}
and the biorthogonality relation leads to the normalization of the right eigenvectors as
\begin{align}
\left<\bm{r}_{q}|\mathcal{F}^{2}|\bm{r}_{q'}\right>=\delta_{qq'}.\label{EqS14}
\end{align}

\section{Berry phase}

The calculation of the Berry phase follows the standard derivation~\cite{Berry1984_supp}, apart from a slight complication in the formalism due to the time-evolution operator being quasi-Hermitian. Equation~\eqref{EqS8} may be rewritten as
\begin{align}
\textrm{d}_{t}\left|\bm{v}\right>=\mathcal{D}_{\bm{k}}\left|\bm{v}\right>,\label{EqS15}
\end{align}
where the matrix $\mathcal{D}$ is now indexed with a parameter $\bm{k}$. In practice, in a translationally invariant system, $\mathcal{D}$ may be reshaped in a block-diagonal form, where each block is indexed by a wave vector $\bm{k}$, the possible values of which are determined by the system size and geometry. Assume that $\bm{k}$ evolves in time according to a predefined function, i.e., a path is mapped out in the Brillouin zone. We assume that the state $\left|\bm{v}\right>$ is initialized in a right eigenstate of the system $\left|\bm{r}_{n,\bm{k}\left(t=0\right)}\right>$. In the adiabatic approach, we approximate the time evolution defined in Eq.~\eqref{EqS15} by assuming that the system stays in an eigenstate,
\begin{align}
\left|\bm{v}\right>\left(t\right)\approx c_{n}\left(t\right)\textrm{e}^{\int\mathcal{D}_{\bm{k}\left(t'\right)}\textrm{d}t'}\left|\bm{r}_{n,\bm{k}\left(t\right)}\right>.\label{EqS16}
\end{align}
Substituting Eq.~\eqref{EqS16} into Eq.~\eqref{EqS15} and using the biorthogonality relation, the phase factor $c_{n}\left(t\right)$ is explicitly given as
\begin{align}
c_{n}\left(t\right)&=\textrm{e}^{-\int\left<\bm{l}_{n,\bm{k}\left(t'\right)}\right|\left(\textrm{d}_{t}\left|\bm{r}_{n,\bm{k}\left(t'\right)}\right>\right)\textrm{d}t'}\nonumber\\&=\textrm{e}^{-\int\left<\bm{l}_{n,\bm{k}}\right|\left(\partial_{\bm{k}}\left|\bm{r}_{n,\bm{k}}\right>\right)\textrm{d}\bm{k}}.\label{EqS17}
\end{align}
On the right-hand side, the phase is expressed by a line integral in parameter space, no longer requiring any assumptions about the time evolution. Due to the non-trivial topology of the parameter space, being a torus in the case of the Brillouin zone, the Berry phase factor $c_{n}$ does not vanish over a closed curve. It may also be characterized by the Berry connection
\begin{align}
A_{n}^{\alpha}\left(\bm{k}\right)=\textrm{i}\left<\bm{l}_{n,\bm{k}}\right|\left(\partial_{k^{\alpha}}\left|\bm{r}_{n,\bm{k}}\right>\right),\label{EqS18}
\end{align}
and the Berry curvature, which in a three-dimensional parameter space reads
\begin{align}
    B^{\alpha}_{n}\left(\bm{k}\right) =
    {\rm i}\epsilon^{\alpha\beta\gamma}\left(\partial_{k^{\beta}}\left<\bm{l}_{n,\bm{k}}\right|\right)\left(\partial_{k^{\gamma}}\left|\bm{r}_{n,\bm{k}}\right>\right).\label{EqS19}
\end{align}
In the two-dimensional Brillouin zone in the $xy$ plane considered in the main text, only the $z$ component of the Berry curvature remains. Using the eigenvalue equation (2) and the biorthogonality relations yields
\begin{align}
\left(\omega_{n}-\omega_{m}\right)\left<\bm{l}_{m,\bm{k}}\right|\left(\partial_{k^{\alpha}}\left|\bm{r}_{n,\bm{k}}\right>\right)=\left<\bm{l}_{m,\bm{k}}|\partial_{k^{\alpha}}\mathcal{D}_{\bm{k}}|\bm{r}_{n,\bm{k}}\right>\label{EqS20}
\end{align}
for $m\neq n$, which leads to the alternative expression Eq.~(12) in the main text. This has the advantage that the derivatives of the matrix $\mathcal{D}_{\bm{k}}$ may often be calculated analytically, while calculating the derivatives of the eigenvectors numerically is difficult, because their phase may vary between different wave vectors. An alternative form of the Berry curvature may be derived by expressing the left eigenvectors via Eq.~\eqref{EqS13}, substituting the expression $\left|\bm{r}_{n,\bm{k}}\right>=\left[\bm{v}_{n,\bm{k}},\omega_{n,\bm{k}}\bm{v}_{n,\bm{k}}\right]$ enforced by the matrix structure, and calculating the derivatives of $\mathcal{D}_{\bm{k}}$ where only $\mathcal{H}_{\textrm{SW},\bm{k}}$ depends on the wave vector. This results in
\begin{align}
B^{\alpha}_{n}\left(\bm{k}\right) =&
    {\rm i}\epsilon^{\alpha\beta\gamma}\sum_{m\neq n}\omega_{n,\bm{k}}\omega_{m,\bm{k}}\times\\\nonumber&\frac{\left<\bm{v}_{n,\bm{k}}|\partial_{k^{\beta}}\mathcal{H}_{\textrm{SW},\bm{k}}|\bm{v}_{m,\bm{k}}\right>\left<\bm{v}_{m,\bm{k}}|\partial_{k^{\gamma}}\mathcal{H}_{\textrm{SW},\bm{k}}|\bm{v}_{n,\bm{k}}\right>}{\left(\omega_{n,\bm{k}}-\omega_{m,\bm{k}}\right)^2}.\label{EqS21}
\end{align}
This is similar to the formula in Ref.~\cite{Zhang2013_supp}, with the caveat that the eigenvectors follow the normalization condition derived from Eq.~\eqref{EqS14},
\begin{align}
\bigg<\bm{v}_{n,\bm{k}}\left|\left[\mathcal{A}\mathcal{B}^{-\frac{1}{2}}\right]^{2}+\mathcal{B}^{-1}\omega^{2}_{n,\bm{k}}
\right|\bm{v}_{n',\bm{k}'}\bigg>=\delta_{nn'}\delta_{\bm{k}\bm{k}'}.
\end{align}
Alternatively, the Berry curvature may also be written as proposed in Ref.~\cite{Shindou2013_supp},
\begin{align}
    B^{\alpha}_{n}\left(\bm{k}\right) =
    {\rm i}\epsilon^{\alpha\beta\gamma}\textrm{Tr}\left[\left(\mathbb{1}_{\bm{k}}-P_{n,\bm{k}}\right)\partial_{k^{\beta}}P_{n,\bm{k}}\partial_{k^{\gamma}}P_{n,\bm{k}}\right],\label{EqS23}
\end{align}
where $P_{n,\bm{k}}=\left|\bm{r}_{n,\bm{k}}\right>\left<\bm{l}_{n,\bm{k}}\right|$ is the (non-orthogonal) projection onto the subspace of the eigenstate.

\section{Gap scaling with the pseudodipolar interaction}

In the absence of spin inertia, the gap in the magnon spectrum of the ferromagnet on the honeycomb lattice is opened in the $K$ and $K'$ points. In the $K=\frac{2\pi}{a}\left(0,\frac{2}{3\sqrt{3}}\right)$ point, the spin-wave Hamiltonian from Eq.~(5) in the main text reads
 \begin{align}
    \mathcal{H}_{\textrm{SW},K}=& \begin{pmatrix}
         -3J^{zz} & 0& 0 &\frac{3}{2}F\\[10pt]
         0 & -3J^{zz} & 0 & 0\\[10pt]
         0 &0 & -3J^{zz} & 0\\[10pt]
         \frac{3}{2}F & 0 &0 & -3J^{zz}
      \end{pmatrix}\,,
\end{align}
The magnon frequencies are given by the eigenvalues of the matrix $\bm{\sigma}^{z}\mathcal{H}_{\textrm{SW}}$. Treating the pseudodipolar interaction as a perturbation, the unperturbed frequencies are $\mp 3J^{zz}$, both of them twice degenerate and of counterclockwise polarization. In first order of the perturbation theory, the matrix elements of the perturbation matrix proportional to $F$ have to be calculated on the degenerate subspaces. However, since these vectors only have matrix elements either in the upper left or the lower right block of the $4\times 4$ matrix, and the perturbation matrix is offdiagonal in the blocks (connecting counterclockwise modes with clockwise modes), these terms all vanish. The second-order corrections in perturbation theory will be finite, meaning that the gap scales with $\left|F\right|^{2}$. This is different from the case of the next-nearest neighbor Dzyaloshinsky-Moriya interaction~\cite{mcclarty2022_supp}, which gives nonvanishing contributions to the diagonal terms of the matrix in the $K$ point, thereby opening a gap in first order of the perturbation theory.

In the presence of spin inertia, the precessional and nutational bands may become degenerate along a general contour in the Brillouin zone. Restricting the eigenvectors $\left|\bm{r}_{\bm{k}}\right>$ to the spin components $\bm{v}_{\bm{k}}$, one of the degenerate eigenstates for $F=0$ is restricted to the first two components (counterclockwise mode), while the other one is restricted to the second two components (clockwise mode). The pseudodipolar term breaking the conservation of total angular momentum introduces a matrix element between the counterclockwise and the clockwise mode already in first order of perturbation theory, thereby opening a gap which scales as $\left|F\right|$.

\section{Topological phase transition in the honeycomb lattice}

\begin{figure*}
\includegraphics[width=2\columnwidth]{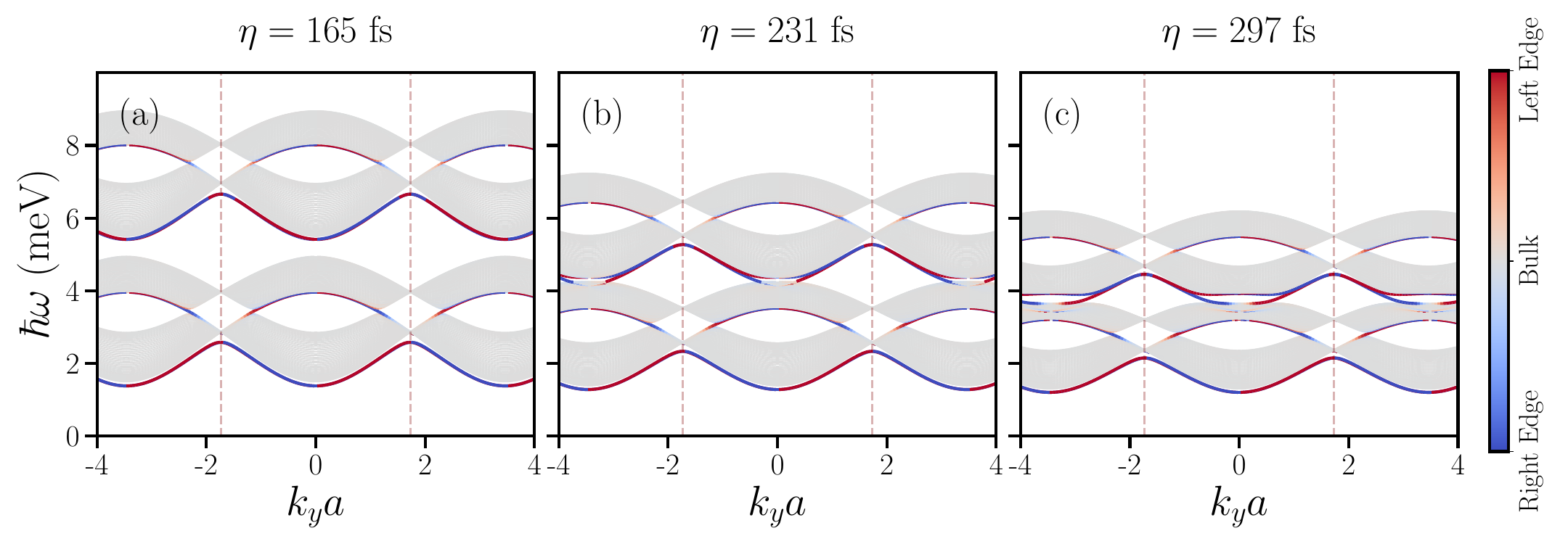}
    \caption{Magnon band structure in a slab geometry. The value of $\eta$ is indicated in each panel, while the other parameters are the same as in Table~I in the main text. The system is $40$ unit cells wide along the $x$ direction with open boundary conditions, while it is infinite along the $y$ direction with periodic boundary conditions, meaning that the eigenstates can be indexed by the wave vector component $k_{y}$. $a$ denotes the distance between nearest-neighbor atoms in the lattice, i.e., the lattice constant is $\sqrt{3}a$. Dotted lines show the boundaries of the one-dimensional Brillouin zone. Coloring shows the localization defined in Eq.~(11) in the main text. }
\label{figS1}
 \end{figure*}

\begin{figure*}
     \includegraphics[width=2\columnwidth]{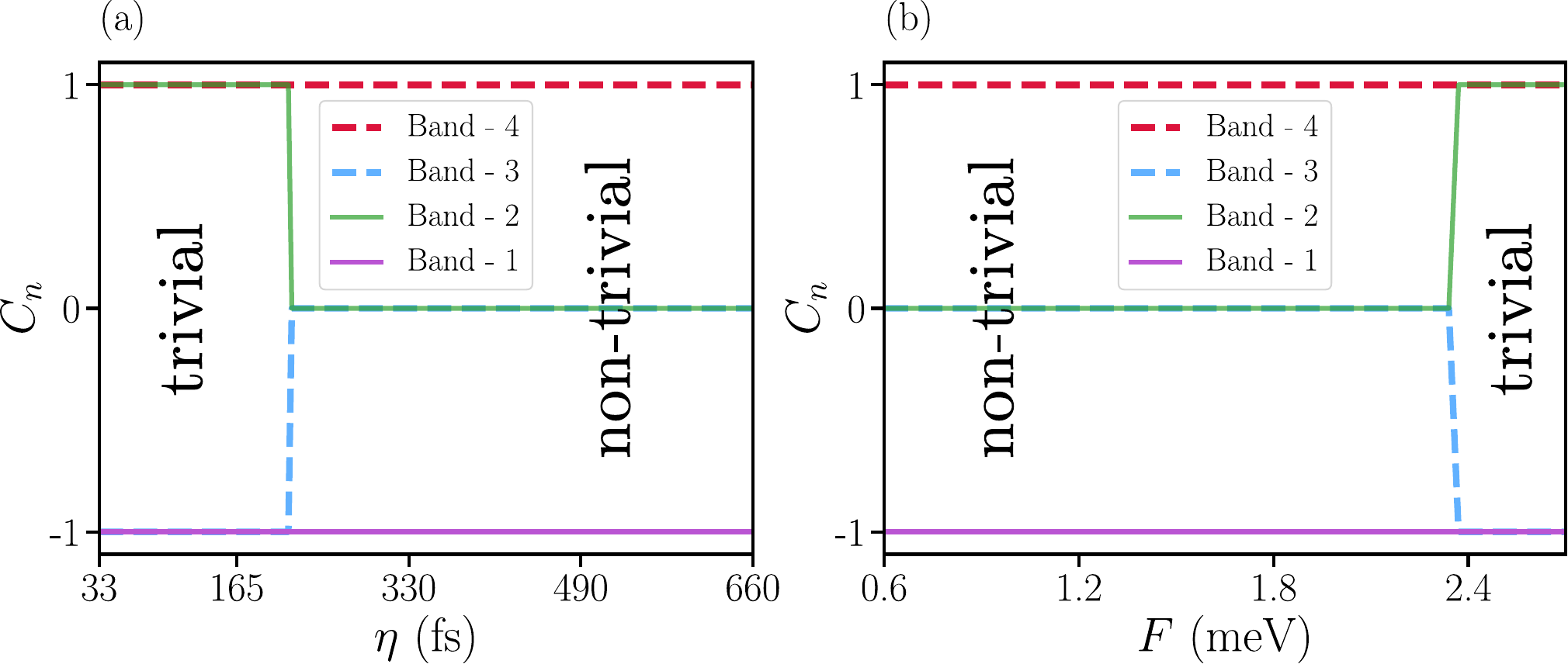}
     \caption{Chern numbers of the magnon bands. The parameters were the same as in Table~I in the main text, except from the inertial relaxation time $\eta$ in panel (a) and the pseudodipolar interaction $F$ in panel (b) which were varied as indicated. Trivial and non-trivial denote the topological character of the gap between the second and third bands.}\label{figS2}
 \end{figure*}

As described in the main text, two different types of gaps may be identified in the magnon spectrum of the ferromagnetic honeycomb lattice. The gap inside the precessional and nutational bands from the Dirac cones at the $K$ and $K'$ points opens for arbitrarily small values of the pseudo-dipolar interaction $F$, and is always topologically non-trivial. However, the precessional and nutational bands are separated by a gap for low values of the inertial relaxation time $\eta$, as illustrated in Fig.~\ref{figS1}(a). Edge states are visible in the slab geometry below the nutational bands, but similarly to the edge states below the precessional bands, these bands do not connect two bands, and are of topologically trivial origin. As $\eta$ is increased, the gap between the bands closes and reopens. As shown in Fig.~\ref{figS1}(b), after the reopening the edge states connect the precessional to the nutational band. Increasing the inertial relaxation time further in Fig.~\ref{figS1}(c) increases the gap size, despite the general reduction in the band widths with increasing $\eta$. This topological phase transition happens close to
$\eta=230$~fs, as can be seen from the change in the Chern numbers in Fig.~\ref{figS2}(a). For lower values of $\eta$, the Chern numbers of the bands are alternating between $-1$ and $1$, meaning that their sum for the first two bands is zero, indicating no topologically non-trivial edge state between the second and third bands. As $\eta$ is increased, the Chern numbers of the second and third bands change to $0$, explaining the formation of edge states between them. The topological character of the bands may also be modified by changing the strength of the pseudodipolar interaction strength, as shown in Fig.~\ref{figS2}(b). For higher values of $F$, the width of the bands decreases, which increases the distance between the precessional and nutational bands, and transforms the gap between the second and third bands back to topologically trivial where the Chern numbers in all bands are $+1$ or $-1$.

\section{Magnon Chern insulator in a single-sublattice ferromagnet}

Here, we consider a two-dimensional ferromagnet on a triangular lattice, which commonly occurs in epitaxially grown transition-metal layers on nonmagnetic substrates such as Co on Au(111)~\cite{Pommier1990_supp,Allenspach1990_supp}. The substrate breaks the inversion symmetry of the structure, leaving a $C_{3\textrm{v}}$ point-group symmetry with the direction perpendicular to the surface denoted by $\bm{\hat{z}}$. Restricting the model to nearest-neighbor pair interactions, it can be described by the Hamiltonian
\begin{align}
 \mathcal{H}  = &\frac{1}{2}\sum_{\langle i, j\rangle} \left[J\bm{S}_{i} \bm{S}_{j}-\Delta J^{zz}\left(2S_{i}^{z}S_{j}^{z}-S_{i}^{x}S_{j}^{x}-S_{i}^{y}S_{j}^{y}\right) \right.\nonumber\\&\left.+F \left\{\left(\bm{S}_{i}\hat{\bm{r}}_{ij}\right)\left(\bm{S}_{j} \hat{\bm{r}}_{ij}\right)-\left[\bm{S}_{i}\left(\hat{\bm{z}}\times\hat{\bm{r}}_{ij}\right)\right]\left[\bm{S}_{j}\left(\hat{\bm{z}}\times \hat{\bm{r}}_{ij}\right)\right]\right\}\right.\nonumber\\&\left.+\left(D_{ij}^{z}\hat{\bm{z}}+D^{\parallel}\hat{\bm{z}}\times\hat{\bm{r}}_{ij}\right)\left(\bm{S}_{i}\times\bm{S}_{j}\right)\right.\nonumber\\&\left.+\Delta J^{\bot}\left\{\left[\bm{S}_{i}\left(\hat{\bm{z}}\times\hat{\bm{r}}_{ij}\right)\right]S_{j}^{z}+S_{i}^{z}\left[\bm{S}_{j}\left(\hat{\bm{z}}\times\hat{\bm{r}}_{ij}\right)\right]\right\}\right]\nonumber\\&+\sum_{i}\left[K\left(S_{i}^{z}\right)^{2}+MB^{z}S_{i}^{z}\right]\,,\label{EqS24}
\end{align}
where $J$ is the exchange interaction, $\Delta J^{zz}$ is the uniaxial two-site anisotropy, $F$ is the pseudodipolar interaction, $D_{ij}^{z}$ and $D^{\parallel}$ are out-of-plane and in-plane components of the Dzyaloshinsky-Moriya interaction, respectively (note that $D_{ij}^{z}$ has opposite signs for opposite directions of $\hat{\bm{r}}_{ij}$), $\Delta J^{\bot}$ is an anisotropy term connecting the in-plane and out-of-plane components, $K$ is the uniaxial single-site anisotropy, and $B^{z}$ is the external field applied along the symmetry axis. This includes all possible coefficients allowed by symmetry considerations.

Assuming a ferromagnetic ground state along the $\hat{\bm{z}}$ direction, the spin-wave Hamiltonian in Fourier space may be written as
\begin{align}
\mathcal{H}_{\textrm{SW},\bm{k}}=\begin{bmatrix}D_{0,\bm{k}}+D_{\textrm{nr},\bm{k}}&D_{\textrm{a},\bm{k}}\\D^{*}_{\textrm{a},\bm{k}}&D_{0,\bm{k}}-D_{\textrm{nr},\bm{k}}\end{bmatrix},\label{EqS25}
\end{align}
with
\begin{align}
D_{0,\bm{k}}=&2\left(J+\Delta J^{zz}\right)\left[\cos{\left(k_{x}a\right)}+2\cos{\left(\frac{k_{x}}{2}a\right)}\cos{\left(\frac{\sqrt{3}k_{y}}{2}a\right)}\right]\nonumber\\&-6\left(J-2\Delta J^{zz}\right)-2K-MB^{z},\label{EqS26}\\
D_{\textrm{nr},\bm{k}}=&-2D^{z}\left[\sin{\left(k_{x}a\right)}-2\sin{\left(\frac{k_{x}}{2}a\right)}\cos{\left(\frac{\sqrt{3}k_{y}}{2}a\right)}\right],\label{EqS27}\\
D_{\textrm{a},\bm{k}}=&2F\left[\cos{\left(k_{x}a\right)}-\cos{\left(\frac{k_{x}}{2}a\right)}\cos{\left(\frac{\sqrt{3}k_{y}}{2}a\right)}\right]\nonumber\\&-\textrm{i}2\sqrt{3}F\sin{\left(\frac{k_{x}}{2}a\right)}\sin{\left(\frac{\sqrt{3}k_{y}}{2}a\right)},\label{EqS28}
\end{align}
where $D_{0,\bm{k}},D_{\textrm{nr},\bm{k}}$, and $D_{\textrm{a},\bm{k}}$ are the normal (even in $\bm{k}$), nonreciprocal (odd in $\bm{k}$), and anomalous (pairing $\bm{S}^{+}$ with $\bm{S}^{+}$ and $\bm{S}^{-}$ with $\bm{S}^{-}$) terms, respectively. Note that the $D^{\parallel}$ and $\Delta J^{\bot}$ parameters do not influence the spectrum in linear spin-wave theory for this magnetization orientation, and the role of the two-site and single-site anisotropies as well as that of the external field are mostly interchangeable.

The magnon frequencies may be determined by solving Eq.~(3) in the main text, which may be rearranged as
\begin{align}
&\left(-\omega_{n,\bm{k}}^{2}\eta+\frac{\gamma}{M}D_{0,\bm{k}}\right)^{2}-\left(\omega_{n,\bm{k}}+\frac{\gamma}{M}D_{\textrm{nr},\bm{k}}\right)^{2}\nonumber\\&-\left|\frac{\gamma}{M}D_{\textrm{a},\bm{k}}\right|^{2}=0.\label{EqS29}
\end{align}
In the absence of inertia ($\eta=0$), the single pair of solutions $\omega_{1/2,\bm{k}}=\pm\omega_{\bm{k}}$ form particle-hole partners, which cannot be degenerate apart from the trivial case where $\mathcal{H}_{\textrm{SW},\bm{k}}=\mathbb{0}$. It is not possible to form topologically non-trivial bands in this case, since that would require changing the sign of one of the eigenvalues inside the Brillouin zone, which would destabilize the ferromagnetic ground state.

When spin inertia is taken into account, there are two pairs of particle-hole partners. The positive-frequency solutions may be approximately described by a two-level Hamiltonian
\begin{align}
\mathcal{H}_{2}=d_{0,\bm{k}}\mathbb{1}+d_{\alpha,\bm{k}}\sigma^{\alpha},\label{EqS30}
\end{align}
where $d_{0,\bm{k}}$ is the mean of the two eigenfrequencies, $d_{x,\bm{k}}=\frac{\gamma}{M}\textrm{Re}D_{\textrm{a},\bm{k}}$, $d_{y,\bm{k}}=-\frac{\gamma}{M}\textrm{Im}D_{\textrm{a},\bm{k}}$, and $d_{x,\bm{k}}=\frac{\gamma}{M}D_{0,\bm{k}}-\eta\left(\frac{\gamma}{M}D_{\textrm{nr},\bm{k}}\right)^{2}$. The two solutions are degenerate when $d_{x,\bm{k}}=d_{y,\bm{k}}=d_{z,\bm{k}}=0$; note that this condition is exact also in Eq.~\eqref{EqS29}, not only on the level of the two-level approximation. Since there are three conditions and the parameter space is two-dimensional, such a degeneracy only occurs for specific model parameter values where the gap closes.

The topology of the band structure may be characterized by how many times the direction of the vector $\left(d_{x,\bm{k}},d_{y,\bm{k}},d_{z,\bm{k}}\right)$ wraps the two-dimensional unit sphere. This corresponds to the Chern number of the considered bands up to a sign, with the two bands having opposite Chern numbers. The Chern number may only be nonzero if all three components change sign in the Brillouin zone. While this is always true for $d_{x,\bm{k}}$ and $d_{y,\bm{k}}$, $d_{z,\bm{k}}$ is usually positive. Typically the exchange interaction is stronger than the Dzyaloshinsky-Moriya interaction, $D_{0,\bm{k}}>\left|D_{\textrm{nr},\bm{k}}\right|$, and $D_{0,\bm{k}}$ is required to be positive everywhere in the Brillouin zone in order for the ground state to be stable. $D_{0,\bm{k}}$ is smallest in the vicinity of $\bm{k}=\bm{0}$, but at this point $\left|D_{\textrm{nr},\bm{k}}\right|$ also vanishes since it is antisymmetric in $\bm{k}$. Furthermore, the parameter $\eta\frac{\gamma}{M}\left|D_{\textrm{nr},\bm{k}}\right|$ is often small since it scales with the strength of the Dzyaloshinsky-Moriya interaction and the inertial relaxation time, both of which emerge due to the spin--orbit coupling. While mathematically it is possible to select such a high value of $\eta$ that a sign change occurs and the Chern number becomes finite, this may be difficult to observe in actual materials.

Compared to the honeycomb lattice discussed in the main text, one can see that the pseudodipolar interaction $F$ is required for the formation of the topologically non-trivial phase in both cases, since this term breaks the conservation of angular momentum. While the Dzyaloshinsky-Moriya interaction played an unimportant role for the honeycomb lattice, it must assume very high values in the triangular lattice, possibly becoming stronger than the exchange interaction. In contrast, for the honeycomb lattice the distance between the precessional and nutational bands may be decreased by increasing the dimensionless parameter $\eta\gamma M^{-1}\left|J\right|$, i.e., for higher values of either the inertial relaxation time or the exchange interaction.


\begin{thebibliography}{50}%
\setcounter{NAT@ctr}{0}
\makeatletter
\providecommand \@ifxundefined [1]{%
 \@ifx{#1\undefined}
}%
\providecommand \@ifnum [1]{%
 \ifnum #1\expandafter \@firstoftwo
 \else \expandafter \@secondoftwo
 \fi
}%
\providecommand \@ifx [1]{%
 \ifx #1\expandafter \@firstoftwo
 \else \expandafter \@secondoftwo
 \fi
}%
\providecommand \natexlab [1]{#1}%
\providecommand \enquote  [1]{``#1''}%
\providecommand \bibnamefont  [1]{#1}%
\providecommand \bibfnamefont [1]{#1}%
\providecommand \citenamefont [1]{#1}%
\providecommand \href@noop [0]{\@secondoftwo}%
\providecommand \href [0]{\begingroup \@sanitize@url \@href}%
\providecommand \@href[1]{\@@startlink{#1}\@@href}%
\providecommand \@@href[1]{\endgroup#1\@@endlink}%
\providecommand \@sanitize@url [0]{\catcode `\\12\catcode `\$12\catcode
  `\&12\catcode `\#12\catcode `\^12\catcode `\_12\catcode `\%12\relax}%
\providecommand \@@startlink[1]{}%
\providecommand \@@endlink[0]{}%
\providecommand \url  [0]{\begingroup\@sanitize@url \@url }%
\providecommand \@url [1]{\endgroup\@href {#1}{\urlprefix }}%
\providecommand \urlprefix  [0]{URL }%
\providecommand \Eprint [0]{\href }%
\providecommand \doibase [0]{https://doi.org/}%
\providecommand \selectlanguage [0]{\@gobble}%
\providecommand \bibinfo  [0]{\@secondoftwo}%
\providecommand \bibfield  [0]{\@secondoftwo}%
\providecommand \translation [1]{[#1]}%
\providecommand \BibitemOpen [0]{}%
\providecommand \bibitemStop [0]{}%
\providecommand \bibitemNoStop [0]{.\EOS\space}%
\providecommand \EOS [0]{\spacefactor3000\relax}%
\providecommand \BibitemShut  [1]{\csname bibitem#1\endcsname}%
\let\auto@bib@innerbib\@empty
\bibitem [{\citenamefont {Siegmann}\ \emph {et~al.}(1995)\citenamefont
  {Siegmann}, \citenamefont {Garwin}, \citenamefont {Prescott}, \citenamefont
  {Heidmann}, \citenamefont {Mauri}, \citenamefont {Weller}, \citenamefont
  {Allenspach},\ and\ \citenamefont {Weber}}]{SIEGMANN1995L8}%
  \BibitemOpen
  \bibfield  {author} {\bibinfo {author} {\bibfnamefont {H.~C.}\ \bibnamefont
  {Siegmann}}, \bibinfo {author} {\bibfnamefont {E.~L.}\ \bibnamefont
  {Garwin}}, \bibinfo {author} {\bibfnamefont {C.~Y.}\ \bibnamefont
  {Prescott}}, \bibinfo {author} {\bibfnamefont {J.}~\bibnamefont {Heidmann}},
  \bibinfo {author} {\bibfnamefont {D.}~\bibnamefont {Mauri}}, \bibinfo
  {author} {\bibfnamefont {D.}~\bibnamefont {Weller}}, \bibinfo {author}
  {\bibfnamefont {R.}~\bibnamefont {Allenspach}},\ and\ \bibinfo {author}
  {\bibfnamefont {W.}~\bibnamefont {Weber}},\ }\bibfield  {title} {\bibinfo
  {title} {Magnetism with picosecond field pulses},\ }\href
  {https://doi.org/http://dx.doi.org/10.1016/0304-8853(95)00602-8} {\bibfield
  {journal} {\bibinfo  {journal} {J. Magn. Magn. Mater.}\ }\textbf {\bibinfo
  {volume} {151}},\ \bibinfo {pages} {L8 } (\bibinfo {year}
  {1995})}\BibitemShut {NoStop}%
\bibitem [{\citenamefont {Beaurepaire}\ \emph {et~al.}(1996)\citenamefont
  {Beaurepaire}, \citenamefont {Merle}, \citenamefont {Daunois},\ and\
  \citenamefont {Bigot}}]{Beaurepaire1996}%
  \BibitemOpen
  \bibfield  {author} {\bibinfo {author} {\bibfnamefont {E.}~\bibnamefont
  {Beaurepaire}}, \bibinfo {author} {\bibfnamefont {J.-C.}\ \bibnamefont
  {Merle}}, \bibinfo {author} {\bibfnamefont {A.}~\bibnamefont {Daunois}},\
  and\ \bibinfo {author} {\bibfnamefont {J.-Y.}\ \bibnamefont {Bigot}},\
  }\bibfield  {title} {\bibinfo {title} {Ultrafast spin dynamics in
  ferromagnetic nickel},\ }\href {https://doi.org/10.1103/PhysRevLett.76.4250}
  {\bibfield  {journal} {\bibinfo  {journal} {Phys. Rev. Lett.}\ }\textbf
  {\bibinfo {volume} {76}},\ \bibinfo {pages} {4250} (\bibinfo {year}
  {1996})}\BibitemShut {NoStop}%
\bibitem [{\citenamefont {Kampfrath}\ \emph {et~al.}(2023)\citenamefont
  {Kampfrath}, \citenamefont {Kirilyuk}, \citenamefont {Mangin}, \citenamefont
  {Sharma},\ and\ \citenamefont {Weinelt}}]{Kampfrath2023}%
  \BibitemOpen
  \bibfield  {author} {\bibinfo {author} {\bibfnamefont {T.}~\bibnamefont
  {Kampfrath}}, \bibinfo {author} {\bibfnamefont {A.}~\bibnamefont {Kirilyuk}},
  \bibinfo {author} {\bibfnamefont {S.}~\bibnamefont {Mangin}}, \bibinfo
  {author} {\bibfnamefont {S.}~\bibnamefont {Sharma}},\ and\ \bibinfo {author}
  {\bibfnamefont {M.}~\bibnamefont {Weinelt}},\ }\bibfield  {title} {\bibinfo
  {title} {Ultrafast and terahertz spintronics: Guest editorial},\ }\href
  {https://doi.org/10.1063/5.0167151} {\bibfield  {journal} {\bibinfo
  {journal} {Applied Physics Letters}\ }\textbf {\bibinfo {volume} {123}},\
  \bibinfo {pages} {050401} (\bibinfo {year} {2023})}\BibitemShut {NoStop}%
\bibitem [{\citenamefont {Ciornei}\ \emph {et~al.}(2011)\citenamefont
  {Ciornei}, \citenamefont {Rub\'{\i}},\ and\ \citenamefont
  {Wegrowe}}]{Ciornei2011}%
  \BibitemOpen
  \bibfield  {author} {\bibinfo {author} {\bibfnamefont {M.-C.}\ \bibnamefont
  {Ciornei}}, \bibinfo {author} {\bibfnamefont {J.~M.}\ \bibnamefont
  {Rub\'{\i}}},\ and\ \bibinfo {author} {\bibfnamefont {J.-E.}\ \bibnamefont
  {Wegrowe}},\ }\bibfield  {title} {\bibinfo {title} {Magnetization dynamics in
  the inertial regime: Nutation predicted at short time scales},\ }\href
  {https://doi.org/10.1103/PhysRevB.83.020410} {\bibfield  {journal} {\bibinfo
  {journal} {Phys. Rev. B}\ }\textbf {\bibinfo {volume} {83}},\ \bibinfo
  {pages} {020410} (\bibinfo {year} {2011})}\BibitemShut {NoStop}%
\bibitem [{\citenamefont {Suhl}(1998)}]{Suhl1998}%
  \BibitemOpen
  \bibfield  {author} {\bibinfo {author} {\bibfnamefont {H.}~\bibnamefont
  {Suhl}},\ }\bibfield  {title} {\bibinfo {title} {Theory of the magnetic
  damping constant},\ }\href {https://doi.org/10.1109/20.706720} {\bibfield
  {journal} {\bibinfo  {journal} {IEEE Trans. Magn.}\ }\textbf {\bibinfo
  {volume} {34}},\ \bibinfo {pages} {1834} (\bibinfo {year}
  {1998})}\BibitemShut {NoStop}%
\bibitem [{\citenamefont {Mondal}\ \emph {et~al.}(2017)\citenamefont {Mondal},
  \citenamefont {Berritta}, \citenamefont {Nandy},\ and\ \citenamefont
  {Oppeneer}}]{Mondal2017Nutation}%
  \BibitemOpen
  \bibfield  {author} {\bibinfo {author} {\bibfnamefont {R.}~\bibnamefont
  {Mondal}}, \bibinfo {author} {\bibfnamefont {M.}~\bibnamefont {Berritta}},
  \bibinfo {author} {\bibfnamefont {A.~K.}\ \bibnamefont {Nandy}},\ and\
  \bibinfo {author} {\bibfnamefont {P.~M.}\ \bibnamefont {Oppeneer}},\
  }\bibfield  {title} {\bibinfo {title} {Relativistic theory of magnetic
  inertia in ultrafast spin dynamics},\ }\href
  {https://doi.org/10.1103/PhysRevB.96.024425} {\bibfield  {journal} {\bibinfo
  {journal} {Phys. Rev. B}\ }\textbf {\bibinfo {volume} {96}},\ \bibinfo
  {pages} {024425} (\bibinfo {year} {2017})}\BibitemShut {NoStop}%
\bibitem [{\citenamefont {Olive}\ \emph {et~al.}(2015)\citenamefont {Olive},
  \citenamefont {Lansac}, \citenamefont {Meyer}, \citenamefont {Hayoun},\ and\
  \citenamefont {Wegrowe}}]{Olive2015}%
  \BibitemOpen
  \bibfield  {author} {\bibinfo {author} {\bibfnamefont {E.}~\bibnamefont
  {Olive}}, \bibinfo {author} {\bibfnamefont {Y.}~\bibnamefont {Lansac}},
  \bibinfo {author} {\bibfnamefont {M.}~\bibnamefont {Meyer}}, \bibinfo
  {author} {\bibfnamefont {M.}~\bibnamefont {Hayoun}},\ and\ \bibinfo {author}
  {\bibfnamefont {J.-E.}\ \bibnamefont {Wegrowe}},\ }\bibfield  {title}
  {\bibinfo {title} {{Deviation from the Landau-Lifshitz-Gilbert equation in
  the inertial regime of the magnetization}},\ }\href
  {https://doi.org/10.1063/1.4921908} {\bibfield  {journal} {\bibinfo
  {journal} {J. Appl. Phys.}\ }\textbf {\bibinfo {volume} {117}},\ \bibinfo
  {pages} {213904} (\bibinfo {year} {2015})}\BibitemShut {NoStop}%
\bibitem [{\citenamefont {Mondal}\ \emph {et~al.}(2023)\citenamefont {Mondal},
  \citenamefont {Rózsa}, \citenamefont {Farle}, \citenamefont {Oppeneer},
  \citenamefont {Nowak},\ and\ \citenamefont {Cherkasskii}}]{MONDAL2023170830}%
  \BibitemOpen
  \bibfield  {author} {\bibinfo {author} {\bibfnamefont {R.}~\bibnamefont
  {Mondal}}, \bibinfo {author} {\bibfnamefont {L.}~\bibnamefont {Rózsa}},
  \bibinfo {author} {\bibfnamefont {M.}~\bibnamefont {Farle}}, \bibinfo
  {author} {\bibfnamefont {P.~M.}\ \bibnamefont {Oppeneer}}, \bibinfo {author}
  {\bibfnamefont {U.}~\bibnamefont {Nowak}},\ and\ \bibinfo {author}
  {\bibfnamefont {M.}~\bibnamefont {Cherkasskii}},\ }\bibfield  {title}
  {\bibinfo {title} {Inertial effects in ultrafast spin dynamics},\ }\href
  {https://doi.org/https://doi.org/10.1016/j.jmmm.2023.170830} {\bibfield
  {journal} {\bibinfo  {journal} {Journal of Magnetism and Magnetic Materials}\
  }\textbf {\bibinfo {volume} {579}},\ \bibinfo {pages} {170830} (\bibinfo
  {year} {2023})}\BibitemShut {NoStop}%
\bibitem [{\citenamefont {Olive}\ \emph {et~al.}(2012)\citenamefont {Olive},
  \citenamefont {Lansac},\ and\ \citenamefont {Wegrowe}}]{Olive2012}%
  \BibitemOpen
  \bibfield  {author} {\bibinfo {author} {\bibfnamefont {E.}~\bibnamefont
  {Olive}}, \bibinfo {author} {\bibfnamefont {Y.}~\bibnamefont {Lansac}},\ and\
  \bibinfo {author} {\bibfnamefont {J.-E.}\ \bibnamefont {Wegrowe}},\
  }\bibfield  {title} {\bibinfo {title} {Beyond ferromagnetic resonance: The
  inertial regime of the magnetization},\ }\href
  {https://doi.org/10.1063/1.4712056} {\bibfield  {journal} {\bibinfo
  {journal} {Appl. Phys. Lett.}\ }\textbf {\bibinfo {volume} {100}},\ \bibinfo
  {pages} {192407} (\bibinfo {year} {2012})}\BibitemShut {NoStop}%
\bibitem [{\citenamefont {Neeraj}\ \emph {et~al.}(2021)\citenamefont {Neeraj},
  \citenamefont {Awari}, \citenamefont {Kovalev}, \citenamefont {Polley},
  \citenamefont {Zhou~Hagstr{\"o}m}, \citenamefont {Arekapudi}, \citenamefont
  {Semisalova}, \citenamefont {Lenz}, \citenamefont {Green}, \citenamefont
  {Deinert} \emph {et~al.}}]{neeraj2021inertial}%
  \BibitemOpen
  \bibfield  {author} {\bibinfo {author} {\bibfnamefont {K.}~\bibnamefont
  {Neeraj}}, \bibinfo {author} {\bibfnamefont {N.}~\bibnamefont {Awari}},
  \bibinfo {author} {\bibfnamefont {S.}~\bibnamefont {Kovalev}}, \bibinfo
  {author} {\bibfnamefont {D.}~\bibnamefont {Polley}}, \bibinfo {author}
  {\bibfnamefont {N.}~\bibnamefont {Zhou~Hagstr{\"o}m}}, \bibinfo {author}
  {\bibfnamefont {S.~S. P.~K.}\ \bibnamefont {Arekapudi}}, \bibinfo {author}
  {\bibfnamefont {A.}~\bibnamefont {Semisalova}}, \bibinfo {author}
  {\bibfnamefont {K.}~\bibnamefont {Lenz}}, \bibinfo {author} {\bibfnamefont
  {B.}~\bibnamefont {Green}}, \bibinfo {author} {\bibfnamefont {J.-C.}\
  \bibnamefont {Deinert}}, \emph {et~al.},\ }\bibfield  {title} {\bibinfo
  {title} {Inertial spin dynamics in ferromagnets},\ }\href
  {https://doi.org/10.1038/s41567-020-01040-y} {\bibfield  {journal} {\bibinfo
  {journal} {Nature Physics}\ }\textbf {\bibinfo {volume} {17}},\ \bibinfo
  {pages} {245} (\bibinfo {year} {2021})}\BibitemShut {NoStop}%
\bibitem [{\citenamefont {Unikandanunni}\ \emph {et~al.}(2022)\citenamefont
  {Unikandanunni}, \citenamefont {Medapalli}, \citenamefont {Asa},
  \citenamefont {Albisetti}, \citenamefont {Petti}, \citenamefont {Bertacco},
  \citenamefont {Fullerton},\ and\ \citenamefont
  {Bonetti}}]{unikandanunni2021inertial}%
  \BibitemOpen
  \bibfield  {author} {\bibinfo {author} {\bibfnamefont {V.}~\bibnamefont
  {Unikandanunni}}, \bibinfo {author} {\bibfnamefont {R.}~\bibnamefont
  {Medapalli}}, \bibinfo {author} {\bibfnamefont {M.}~\bibnamefont {Asa}},
  \bibinfo {author} {\bibfnamefont {E.}~\bibnamefont {Albisetti}}, \bibinfo
  {author} {\bibfnamefont {D.}~\bibnamefont {Petti}}, \bibinfo {author}
  {\bibfnamefont {R.}~\bibnamefont {Bertacco}}, \bibinfo {author}
  {\bibfnamefont {E.~E.}\ \bibnamefont {Fullerton}},\ and\ \bibinfo {author}
  {\bibfnamefont {S.}~\bibnamefont {Bonetti}},\ }\bibfield  {title} {\bibinfo
  {title} {Inertial spin dynamics in epitaxial cobalt films},\ }\href
  {https://doi.org/10.1103/PhysRevLett.129.237201} {\bibfield  {journal}
  {\bibinfo  {journal} {Phys. Rev. Lett.}\ }\textbf {\bibinfo {volume} {129}},\
  \bibinfo {pages} {237201} (\bibinfo {year} {2022})}\BibitemShut {NoStop}%
\bibitem [{\citenamefont {De}\ \emph {et~al.}(2025)\citenamefont {De},
  \citenamefont {Schlegel}, \citenamefont {Lentfert}, \citenamefont {Scheuer},
  \citenamefont {Stadtm\"uller}, \citenamefont {Pirro}, \citenamefont {von
  Freymann}, \citenamefont {Nowak},\ and\ \citenamefont
  {Aeschlimann}}]{De2025PRB}%
  \BibitemOpen
  \bibfield  {author} {\bibinfo {author} {\bibfnamefont {A.}~\bibnamefont
  {De}}, \bibinfo {author} {\bibfnamefont {J.}~\bibnamefont {Schlegel}},
  \bibinfo {author} {\bibfnamefont {A.}~\bibnamefont {Lentfert}}, \bibinfo
  {author} {\bibfnamefont {L.}~\bibnamefont {Scheuer}}, \bibinfo {author}
  {\bibfnamefont {B.}~\bibnamefont {Stadtm\"uller}}, \bibinfo {author}
  {\bibfnamefont {P.}~\bibnamefont {Pirro}}, \bibinfo {author} {\bibfnamefont
  {G.}~\bibnamefont {von Freymann}}, \bibinfo {author} {\bibfnamefont
  {U.}~\bibnamefont {Nowak}},\ and\ \bibinfo {author} {\bibfnamefont
  {M.}~\bibnamefont {Aeschlimann}},\ }\bibfield  {title} {\bibinfo {title}
  {Magnetic nutation: Transient separation of magnetization from its angular
  momentum},\ }\href {https://doi.org/10.1103/PhysRevB.111.014432} {\bibfield
  {journal} {\bibinfo  {journal} {Phys. Rev. B}\ }\textbf {\bibinfo {volume}
  {111}},\ \bibinfo {pages} {014432} (\bibinfo {year} {2025})}\BibitemShut
  {NoStop}%
\bibitem [{\citenamefont {Mondal}\ and\ \citenamefont
  {Kamra}(2021)}]{Mondal2021PRBSpinCurrent}%
  \BibitemOpen
  \bibfield  {author} {\bibinfo {author} {\bibfnamefont {R.}~\bibnamefont
  {Mondal}}\ and\ \bibinfo {author} {\bibfnamefont {A.}~\bibnamefont {Kamra}},\
  }\bibfield  {title} {\bibinfo {title} {Spin pumping at terahertz nutation
  resonances},\ }\href {https://doi.org/10.1103/PhysRevB.104.214426} {\bibfield
   {journal} {\bibinfo  {journal} {Phys. Rev. B}\ }\textbf {\bibinfo {volume}
  {104}},\ \bibinfo {pages} {214426} (\bibinfo {year} {2021})}\BibitemShut
  {NoStop}%
\bibitem [{\citenamefont {Winter}\ \emph {et~al.}(2022)\citenamefont {Winter},
  \citenamefont {Gro\ss{}enbach}, \citenamefont {Nowak},\ and\ \citenamefont
  {R\'ozsa}}]{Winter2022}%
  \BibitemOpen
  \bibfield  {author} {\bibinfo {author} {\bibfnamefont {L.}~\bibnamefont
  {Winter}}, \bibinfo {author} {\bibfnamefont {S.}~\bibnamefont
  {Gro\ss{}enbach}}, \bibinfo {author} {\bibfnamefont {U.}~\bibnamefont
  {Nowak}},\ and\ \bibinfo {author} {\bibfnamefont {L.}~\bibnamefont
  {R\'ozsa}},\ }\bibfield  {title} {\bibinfo {title} {Nutational switching in
  ferromagnets and antiferromagnets},\ }\href
  {https://doi.org/10.1103/PhysRevB.106.214403} {\bibfield  {journal} {\bibinfo
   {journal} {Phys. Rev. B}\ }\textbf {\bibinfo {volume} {106}},\ \bibinfo
  {pages} {214403} (\bibinfo {year} {2022})}\BibitemShut {NoStop}%
\bibitem [{\citenamefont {Rodriguez}\ \emph {et~al.}(2024)\citenamefont
  {Rodriguez}, \citenamefont {Cherkasskii}, \citenamefont {Jiang},
  \citenamefont {Mondal}, \citenamefont {Etesamirad}, \citenamefont
  {Tossounian}, \citenamefont {Ivanov},\ and\ \citenamefont
  {Barsukov}}]{Rodriguez2024PRL}%
  \BibitemOpen
  \bibfield  {author} {\bibinfo {author} {\bibfnamefont {R.}~\bibnamefont
  {Rodriguez}}, \bibinfo {author} {\bibfnamefont {M.}~\bibnamefont
  {Cherkasskii}}, \bibinfo {author} {\bibfnamefont {R.}~\bibnamefont {Jiang}},
  \bibinfo {author} {\bibfnamefont {R.}~\bibnamefont {Mondal}}, \bibinfo
  {author} {\bibfnamefont {A.}~\bibnamefont {Etesamirad}}, \bibinfo {author}
  {\bibfnamefont {A.}~\bibnamefont {Tossounian}}, \bibinfo {author}
  {\bibfnamefont {B.~A.}\ \bibnamefont {Ivanov}},\ and\ \bibinfo {author}
  {\bibfnamefont {I.}~\bibnamefont {Barsukov}},\ }\bibfield  {title} {\bibinfo
  {title} {Spin inertia and auto-oscillations in ferromagnets},\ }\href
  {https://doi.org/10.1103/PhysRevLett.132.246701} {\bibfield  {journal}
  {\bibinfo  {journal} {Phys. Rev. Lett.}\ }\textbf {\bibinfo {volume} {132}},\
  \bibinfo {pages} {246701} (\bibinfo {year} {2024})}\BibitemShut {NoStop}%
\bibitem [{\citenamefont {Kikuchi}\ and\ \citenamefont
  {Tatara}(2015)}]{Kikuchi2015}%
  \BibitemOpen
  \bibfield  {author} {\bibinfo {author} {\bibfnamefont {T.}~\bibnamefont
  {Kikuchi}}\ and\ \bibinfo {author} {\bibfnamefont {G.}~\bibnamefont
  {Tatara}},\ }\bibfield  {title} {\bibinfo {title} {Spin dynamics with inertia
  in metallic ferromagnets},\ }\href
  {https://doi.org/10.1103/PhysRevB.92.184410} {\bibfield  {journal} {\bibinfo
  {journal} {Phys. Rev. B}\ }\textbf {\bibinfo {volume} {92}},\ \bibinfo
  {pages} {184410} (\bibinfo {year} {2015})}\BibitemShut {NoStop}%
\bibitem [{\citenamefont {Mondal}\ and\ \citenamefont
  {R\'ozsa}(2022)}]{Mondal2022PRB}%
  \BibitemOpen
  \bibfield  {author} {\bibinfo {author} {\bibfnamefont {R.}~\bibnamefont
  {Mondal}}\ and\ \bibinfo {author} {\bibfnamefont {L.}~\bibnamefont
  {R\'ozsa}},\ }\bibfield  {title} {\bibinfo {title} {Inertial spin waves in
  ferromagnets and antiferromagnets},\ }\href
  {https://doi.org/10.1103/PhysRevB.106.134422} {\bibfield  {journal} {\bibinfo
   {journal} {Phys. Rev. B}\ }\textbf {\bibinfo {volume} {106}},\ \bibinfo
  {pages} {134422} (\bibinfo {year} {2022})}\BibitemShut {NoStop}%
\bibitem [{\citenamefont {Cherkasskii}\ \emph {et~al.}(2024)\citenamefont
  {Cherkasskii}, \citenamefont {Mondal},\ and\ \citenamefont
  {R\'ozsa}}]{CherkasskiiPRB2024}%
  \BibitemOpen
  \bibfield  {author} {\bibinfo {author} {\bibfnamefont {M.}~\bibnamefont
  {Cherkasskii}}, \bibinfo {author} {\bibfnamefont {R.}~\bibnamefont
  {Mondal}},\ and\ \bibinfo {author} {\bibfnamefont {L.}~\bibnamefont
  {R\'ozsa}},\ }\bibfield  {title} {\bibinfo {title} {Inertial spin waves in
  spin spirals},\ }\href {https://doi.org/10.1103/PhysRevB.109.184424}
  {\bibfield  {journal} {\bibinfo  {journal} {Phys. Rev. B}\ }\textbf {\bibinfo
  {volume} {109}},\ \bibinfo {pages} {184424} (\bibinfo {year}
  {2024})}\BibitemShut {NoStop}%
\bibitem [{\citenamefont {Cherkasskii}\ \emph {et~al.}(2021)\citenamefont
  {Cherkasskii}, \citenamefont {Farle},\ and\ \citenamefont
  {Semisalova}}]{Cherkasskii2021}%
  \BibitemOpen
  \bibfield  {author} {\bibinfo {author} {\bibfnamefont {M.}~\bibnamefont
  {Cherkasskii}}, \bibinfo {author} {\bibfnamefont {M.}~\bibnamefont {Farle}},\
  and\ \bibinfo {author} {\bibfnamefont {A.}~\bibnamefont {Semisalova}},\
  }\bibfield  {title} {\bibinfo {title} {Dispersion relation of nutation
  surface spin waves in ferromagnets},\ }\href
  {https://doi.org/10.1103/PhysRevB.103.174435} {\bibfield  {journal} {\bibinfo
   {journal} {Phys. Rev. B}\ }\textbf {\bibinfo {volume} {103}},\ \bibinfo
  {pages} {174435} (\bibinfo {year} {2021})}\BibitemShut {NoStop}%
\bibitem [{\citenamefont {McClarty}(2022)}]{mcclarty2022}%
  \BibitemOpen
  \bibfield  {author} {\bibinfo {author} {\bibfnamefont {P.~A.}\ \bibnamefont
  {McClarty}},\ }\bibfield  {title} {\bibinfo {title} {Topological magnons: A
  review},\ }\href
  {https://doi.org/https://doi.org/10.1146/annurev-conmatphys-031620-104715}
  {\bibfield  {journal} {\bibinfo  {journal} {Annual Review of Condensed Matter
  Physics}\ }\textbf {\bibinfo {volume} {13}},\ \bibinfo {pages} {171}
  (\bibinfo {year} {2022})}\BibitemShut {NoStop}%
\bibitem [{\citenamefont {Zhang}\ \emph {et~al.}(2013)\citenamefont {Zhang},
  \citenamefont {Ren}, \citenamefont {Wang},\ and\ \citenamefont
  {Li}}]{Zhang2013}%
  \BibitemOpen
  \bibfield  {author} {\bibinfo {author} {\bibfnamefont {L.}~\bibnamefont
  {Zhang}}, \bibinfo {author} {\bibfnamefont {J.}~\bibnamefont {Ren}}, \bibinfo
  {author} {\bibfnamefont {J.-S.}\ \bibnamefont {Wang}},\ and\ \bibinfo
  {author} {\bibfnamefont {B.}~\bibnamefont {Li}},\ }\bibfield  {title}
  {\bibinfo {title} {Topological magnon insulator in insulating ferromagnet},\
  }\href {https://doi.org/10.1103/PhysRevB.87.144101} {\bibfield  {journal}
  {\bibinfo  {journal} {Phys. Rev. B}\ }\textbf {\bibinfo {volume} {87}},\
  \bibinfo {pages} {144101} (\bibinfo {year} {2013})}\BibitemShut {NoStop}%
\bibitem [{\citenamefont {Mook}\ \emph
  {et~al.}(2014{\natexlab{a}})\citenamefont {Mook}, \citenamefont {Henk},\ and\
  \citenamefont {Mertig}}]{Mook2014MHE}%
  \BibitemOpen
  \bibfield  {author} {\bibinfo {author} {\bibfnamefont {A.}~\bibnamefont
  {Mook}}, \bibinfo {author} {\bibfnamefont {J.}~\bibnamefont {Henk}},\ and\
  \bibinfo {author} {\bibfnamefont {I.}~\bibnamefont {Mertig}},\ }\bibfield
  {title} {\bibinfo {title} {Magnon hall effect and topology in kagome
  lattices: A theoretical investigation},\ }\href
  {https://doi.org/10.1103/PhysRevB.89.134409} {\bibfield  {journal} {\bibinfo
  {journal} {Phys. Rev. B}\ }\textbf {\bibinfo {volume} {89}},\ \bibinfo
  {pages} {134409} (\bibinfo {year} {2014}{\natexlab{a}})}\BibitemShut
  {NoStop}%
\bibitem [{\citenamefont {Mook}\ \emph
  {et~al.}(2014{\natexlab{b}})\citenamefont {Mook}, \citenamefont {Henk},\ and\
  \citenamefont {Mertig}}]{Mook2014ES}%
  \BibitemOpen
  \bibfield  {author} {\bibinfo {author} {\bibfnamefont {A.}~\bibnamefont
  {Mook}}, \bibinfo {author} {\bibfnamefont {J.}~\bibnamefont {Henk}},\ and\
  \bibinfo {author} {\bibfnamefont {I.}~\bibnamefont {Mertig}},\ }\bibfield
  {title} {\bibinfo {title} {Edge states in topological magnon insulators},\
  }\href {https://doi.org/10.1103/PhysRevB.90.024412} {\bibfield  {journal}
  {\bibinfo  {journal} {Phys. Rev. B}\ }\textbf {\bibinfo {volume} {90}},\
  \bibinfo {pages} {024412} (\bibinfo {year} {2014}{\natexlab{b}})}\BibitemShut
  {NoStop}%
\bibitem [{\citenamefont {Chisnell}\ \emph {et~al.}(2015)\citenamefont
  {Chisnell}, \citenamefont {Helton}, \citenamefont {Freedman}, \citenamefont
  {Singh}, \citenamefont {Bewley}, \citenamefont {Nocera},\ and\ \citenamefont
  {Lee}}]{Chisnell2015}%
  \BibitemOpen
  \bibfield  {author} {\bibinfo {author} {\bibfnamefont {R.}~\bibnamefont
  {Chisnell}}, \bibinfo {author} {\bibfnamefont {J.~S.}\ \bibnamefont
  {Helton}}, \bibinfo {author} {\bibfnamefont {D.~E.}\ \bibnamefont
  {Freedman}}, \bibinfo {author} {\bibfnamefont {D.~K.}\ \bibnamefont {Singh}},
  \bibinfo {author} {\bibfnamefont {R.~I.}\ \bibnamefont {Bewley}}, \bibinfo
  {author} {\bibfnamefont {D.~G.}\ \bibnamefont {Nocera}},\ and\ \bibinfo
  {author} {\bibfnamefont {Y.~S.}\ \bibnamefont {Lee}},\ }\bibfield  {title}
  {\bibinfo {title} {Topological magnon bands in a kagome lattice
  ferromagnet},\ }\href {https://doi.org/10.1103/PhysRevLett.115.147201}
  {\bibfield  {journal} {\bibinfo  {journal} {Phys. Rev. Lett.}\ }\textbf
  {\bibinfo {volume} {115}},\ \bibinfo {pages} {147201} (\bibinfo {year}
  {2015})}\BibitemShut {NoStop}%
\bibitem [{\citenamefont {Shindou}\ \emph {et~al.}(2013)\citenamefont
  {Shindou}, \citenamefont {Matsumoto}, \citenamefont {Murakami},\ and\
  \citenamefont {Ohe}}]{Shindou2013}%
  \BibitemOpen
  \bibfield  {author} {\bibinfo {author} {\bibfnamefont {R.}~\bibnamefont
  {Shindou}}, \bibinfo {author} {\bibfnamefont {R.}~\bibnamefont {Matsumoto}},
  \bibinfo {author} {\bibfnamefont {S.}~\bibnamefont {Murakami}},\ and\
  \bibinfo {author} {\bibfnamefont {J.-i.}\ \bibnamefont {Ohe}},\ }\bibfield
  {title} {\bibinfo {title} {Topological chiral magnonic edge mode in a
  magnonic crystal},\ }\href {https://doi.org/10.1103/PhysRevB.87.174427}
  {\bibfield  {journal} {\bibinfo  {journal} {Phys. Rev. B}\ }\textbf {\bibinfo
  {volume} {87}},\ \bibinfo {pages} {174427} (\bibinfo {year}
  {2013})}\BibitemShut {NoStop}%
\bibitem [{\citenamefont {Roldán-Molina}\ \emph {et~al.}(2016)\citenamefont
  {Roldán-Molina}, \citenamefont {Nunez},\ and\ \citenamefont
  {Fernández-Rossier}}]{RoldnMolina2016}%
  \BibitemOpen
  \bibfield  {author} {\bibinfo {author} {\bibfnamefont {A.}~\bibnamefont
  {Roldán-Molina}}, \bibinfo {author} {\bibfnamefont {A.~S.}\ \bibnamefont
  {Nunez}},\ and\ \bibinfo {author} {\bibfnamefont {J.}~\bibnamefont
  {Fernández-Rossier}},\ }\bibfield  {title} {\bibinfo {title} {Topological
  spin waves in the atomic-scale magnetic skyrmion crystal},\ }\href
  {https://doi.org/10.1088/1367-2630/18/4/045015} {\bibfield  {journal}
  {\bibinfo  {journal} {New Journal of Physics}\ }\textbf {\bibinfo {volume}
  {18}},\ \bibinfo {pages} {045015} (\bibinfo {year} {2016})}\BibitemShut
  {NoStop}%
\bibitem [{\citenamefont {D\'{\i}az}\ \emph {et~al.}(2019)\citenamefont
  {D\'{\i}az}, \citenamefont {Klinovaja},\ and\ \citenamefont
  {Loss}}]{Diaz2019}%
  \BibitemOpen
  \bibfield  {author} {\bibinfo {author} {\bibfnamefont {S.~A.}\ \bibnamefont
  {D\'{\i}az}}, \bibinfo {author} {\bibfnamefont {J.}~\bibnamefont
  {Klinovaja}},\ and\ \bibinfo {author} {\bibfnamefont {D.}~\bibnamefont
  {Loss}},\ }\bibfield  {title} {\bibinfo {title} {{Topological Magnons and
  Edge States in Antiferromagnetic Skyrmion Crystals}},\ }\href
  {https://doi.org/10.1103/PhysRevLett.122.187203} {\bibfield  {journal}
  {\bibinfo  {journal} {Phys. Rev. Lett.}\ }\textbf {\bibinfo {volume} {122}},\
  \bibinfo {pages} {187203} (\bibinfo {year} {2019})}\BibitemShut {NoStop}%
\bibitem [{\citenamefont {Weber}\ \emph {et~al.}(2022)\citenamefont {Weber},
  \citenamefont {Fobes}, \citenamefont {Waizner}, \citenamefont {Steffens},
  \citenamefont {Tucker}, \citenamefont {B\"{o}hm}, \citenamefont {Beddrich},
  \citenamefont {Franz}, \citenamefont {Gabold}, \citenamefont {Bewley},
  \citenamefont {Voneshen}, \citenamefont {Skoulatos}, \citenamefont {Georgii},
  \citenamefont {Ehlers}, \citenamefont {Bauer}, \citenamefont {Pfleiderer},
  \citenamefont {B\"{o}ni}, \citenamefont {Janoschek},\ and\ \citenamefont
  {Garst}}]{Weber2022}%
  \BibitemOpen
  \bibfield  {author} {\bibinfo {author} {\bibfnamefont {T.}~\bibnamefont
  {Weber}}, \bibinfo {author} {\bibfnamefont {D.~M.}\ \bibnamefont {Fobes}},
  \bibinfo {author} {\bibfnamefont {J.}~\bibnamefont {Waizner}}, \bibinfo
  {author} {\bibfnamefont {P.}~\bibnamefont {Steffens}}, \bibinfo {author}
  {\bibfnamefont {G.~S.}\ \bibnamefont {Tucker}}, \bibinfo {author}
  {\bibfnamefont {M.}~\bibnamefont {B\"{o}hm}}, \bibinfo {author}
  {\bibfnamefont {L.}~\bibnamefont {Beddrich}}, \bibinfo {author}
  {\bibfnamefont {C.}~\bibnamefont {Franz}}, \bibinfo {author} {\bibfnamefont
  {H.}~\bibnamefont {Gabold}}, \bibinfo {author} {\bibfnamefont
  {R.}~\bibnamefont {Bewley}}, \bibinfo {author} {\bibfnamefont
  {D.}~\bibnamefont {Voneshen}}, \bibinfo {author} {\bibfnamefont
  {M.}~\bibnamefont {Skoulatos}}, \bibinfo {author} {\bibfnamefont
  {R.}~\bibnamefont {Georgii}}, \bibinfo {author} {\bibfnamefont
  {G.}~\bibnamefont {Ehlers}}, \bibinfo {author} {\bibfnamefont
  {A.}~\bibnamefont {Bauer}}, \bibinfo {author} {\bibfnamefont
  {C.}~\bibnamefont {Pfleiderer}}, \bibinfo {author} {\bibfnamefont
  {P.}~\bibnamefont {B\"{o}ni}}, \bibinfo {author} {\bibfnamefont
  {M.}~\bibnamefont {Janoschek}},\ and\ \bibinfo {author} {\bibfnamefont
  {M.}~\bibnamefont {Garst}},\ }\bibfield  {title} {\bibinfo {title}
  {{Topological magnon band structure of emergent Landau levels in a skyrmion
  lattice}},\ }\href {https://doi.org/10.1126/science.abe4441} {\bibfield
  {journal} {\bibinfo  {journal} {Science}\ }\textbf {\bibinfo {volume}
  {375}},\ \bibinfo {pages} {1025–1030} (\bibinfo {year} {2022})}\BibitemShut
  {NoStop}%
\bibitem [{\citenamefont {Li}\ \emph {et~al.}(2016)\citenamefont {Li},
  \citenamefont {Li}, \citenamefont {Kim}, \citenamefont {Balents},
  \citenamefont {Yu},\ and\ \citenamefont {Chen}}]{Li2016}%
  \BibitemOpen
  \bibfield  {author} {\bibinfo {author} {\bibfnamefont {F.-Y.}\ \bibnamefont
  {Li}}, \bibinfo {author} {\bibfnamefont {Y.-D.}\ \bibnamefont {Li}}, \bibinfo
  {author} {\bibfnamefont {Y.~B.}\ \bibnamefont {Kim}}, \bibinfo {author}
  {\bibfnamefont {L.}~\bibnamefont {Balents}}, \bibinfo {author} {\bibfnamefont
  {Y.}~\bibnamefont {Yu}},\ and\ \bibinfo {author} {\bibfnamefont
  {G.}~\bibnamefont {Chen}},\ }\bibfield  {title} {\bibinfo {title} {Weyl
  magnons in breathing pyrochlore antiferromagnets},\ }\href
  {http://dx.doi.org/10.1038/ncomms12691} {\bibfield  {journal} {\bibinfo
  {journal} {Nature Communications}\ }\textbf {\bibinfo {volume} {7}},\
  \bibinfo {pages} {12691} (\bibinfo {year} {2016})}\BibitemShut {NoStop}%
\bibitem [{\citenamefont {Mook}\ \emph {et~al.}(2016)\citenamefont {Mook},
  \citenamefont {Henk},\ and\ \citenamefont {Mertig}}]{Mook2016}%
  \BibitemOpen
  \bibfield  {author} {\bibinfo {author} {\bibfnamefont {A.}~\bibnamefont
  {Mook}}, \bibinfo {author} {\bibfnamefont {J.}~\bibnamefont {Henk}},\ and\
  \bibinfo {author} {\bibfnamefont {I.}~\bibnamefont {Mertig}},\ }\bibfield
  {title} {\bibinfo {title} {{Tunable Magnon Weyl Points in Ferromagnetic
  Pyrochlores}},\ }\href {https://doi.org/10.1103/PhysRevLett.117.157204}
  {\bibfield  {journal} {\bibinfo  {journal} {Phys. Rev. Lett.}\ }\textbf
  {\bibinfo {volume} {117}},\ \bibinfo {pages} {157204} (\bibinfo {year}
  {2016})}\BibitemShut {NoStop}%
\bibitem [{\citenamefont {Katsura}\ \emph {et~al.}(2010)\citenamefont
  {Katsura}, \citenamefont {Nagaosa},\ and\ \citenamefont {Lee}}]{Katsura2010}%
  \BibitemOpen
  \bibfield  {author} {\bibinfo {author} {\bibfnamefont {H.}~\bibnamefont
  {Katsura}}, \bibinfo {author} {\bibfnamefont {N.}~\bibnamefont {Nagaosa}},\
  and\ \bibinfo {author} {\bibfnamefont {P.~A.}\ \bibnamefont {Lee}},\
  }\bibfield  {title} {\bibinfo {title} {{Theory of the Thermal Hall Effect in
  Quantum Magnets}},\ }\href {https://doi.org/10.1103/PhysRevLett.104.066403}
  {\bibfield  {journal} {\bibinfo  {journal} {Phys. Rev. Lett.}\ }\textbf
  {\bibinfo {volume} {104}},\ \bibinfo {pages} {066403} (\bibinfo {year}
  {2010})}\BibitemShut {NoStop}%
\bibitem [{\citenamefont {Matsumoto}\ and\ \citenamefont
  {Murakami}(2011{\natexlab{a}})}]{Matsumoto2011PRB}%
  \BibitemOpen
  \bibfield  {author} {\bibinfo {author} {\bibfnamefont {R.}~\bibnamefont
  {Matsumoto}}\ and\ \bibinfo {author} {\bibfnamefont {S.}~\bibnamefont
  {Murakami}},\ }\bibfield  {title} {\bibinfo {title} {{Rotational Motion of
  Magnon Wave Packet and Berry Curvature}},\ }\href
  {https://doi.org/10.1103/PhysRevB.84.184406} {\bibfield  {journal} {\bibinfo
  {journal} {Phys. Rev. B}\ }\textbf {\bibinfo {volume} {84}},\ \bibinfo
  {pages} {184406} (\bibinfo {year} {2011}{\natexlab{a}})}\BibitemShut
  {NoStop}%
\bibitem [{\citenamefont {Matsumoto}\ and\ \citenamefont
  {Murakami}(2011{\natexlab{b}})}]{Matsumoto2011PRL}%
  \BibitemOpen
  \bibfield  {author} {\bibinfo {author} {\bibfnamefont {R.}~\bibnamefont
  {Matsumoto}}\ and\ \bibinfo {author} {\bibfnamefont {S.}~\bibnamefont
  {Murakami}},\ }\bibfield  {title} {\bibinfo {title} {{Theoretical Prediction
  of a Rotating Magnon Wave Packet in Ferromagnets}},\ }\href
  {https://doi.org/10.1103/PhysRevLett.106.197202} {\bibfield  {journal}
  {\bibinfo  {journal} {Phys. Rev. Lett.}\ }\textbf {\bibinfo {volume} {106}},\
  \bibinfo {pages} {197202} (\bibinfo {year} {2011}{\natexlab{b}})}\BibitemShut
  {NoStop}%
\bibitem [{\citenamefont {Onose}\ \emph {et~al.}(2010)\citenamefont {Onose},
  \citenamefont {Ideue}, \citenamefont {Katsura}, \citenamefont {Shiomi},
  \citenamefont {Nagaosa},\ and\ \citenamefont {Tokura}}]{Onose2010}%
  \BibitemOpen
  \bibfield  {author} {\bibinfo {author} {\bibfnamefont {Y.}~\bibnamefont
  {Onose}}, \bibinfo {author} {\bibfnamefont {T.}~\bibnamefont {Ideue}},
  \bibinfo {author} {\bibfnamefont {H.}~\bibnamefont {Katsura}}, \bibinfo
  {author} {\bibfnamefont {Y.}~\bibnamefont {Shiomi}}, \bibinfo {author}
  {\bibfnamefont {N.}~\bibnamefont {Nagaosa}},\ and\ \bibinfo {author}
  {\bibfnamefont {Y.}~\bibnamefont {Tokura}},\ }\bibfield  {title} {\bibinfo
  {title} {{Observation of the Magnon Hall Effect}},\ }\href
  {https://doi.org/10.1126/science.1188260} {\bibfield  {journal} {\bibinfo
  {journal} {Science}\ }\textbf {\bibinfo {volume} {329}},\ \bibinfo {pages}
  {297} (\bibinfo {year} {2010})}\BibitemShut {NoStop}%
\bibitem [{\citenamefont {Chen}\ \emph {et~al.}(2018)\citenamefont {Chen},
  \citenamefont {Chung}, \citenamefont {Gao}, \citenamefont {Chen},
  \citenamefont {Stone}, \citenamefont {Kolesnikov}, \citenamefont {Huang},\
  and\ \citenamefont {Dai}}]{Chen2018}%
  \BibitemOpen
  \bibfield  {author} {\bibinfo {author} {\bibfnamefont {L.}~\bibnamefont
  {Chen}}, \bibinfo {author} {\bibfnamefont {J.-H.}\ \bibnamefont {Chung}},
  \bibinfo {author} {\bibfnamefont {B.}~\bibnamefont {Gao}}, \bibinfo {author}
  {\bibfnamefont {T.}~\bibnamefont {Chen}}, \bibinfo {author} {\bibfnamefont
  {M.~B.}\ \bibnamefont {Stone}}, \bibinfo {author} {\bibfnamefont {A.~I.}\
  \bibnamefont {Kolesnikov}}, \bibinfo {author} {\bibfnamefont
  {Q.}~\bibnamefont {Huang}},\ and\ \bibinfo {author} {\bibfnamefont
  {P.}~\bibnamefont {Dai}},\ }\bibfield  {title} {\bibinfo {title}
  {{Topological Spin Excitations in Honeycomb Ferromagnet
  ${\mathrm{CrI}}_{3}$}},\ }\href {https://doi.org/10.1103/PhysRevX.8.041028}
  {\bibfield  {journal} {\bibinfo  {journal} {Phys. Rev. X}\ }\textbf {\bibinfo
  {volume} {8}},\ \bibinfo {pages} {041028} (\bibinfo {year}
  {2018})}\BibitemShut {NoStop}%
\bibitem [{\citenamefont {Chen}\ \emph {et~al.}(2021)\citenamefont {Chen},
  \citenamefont {Chung}, \citenamefont {Stone}, \citenamefont {Kolesnikov},
  \citenamefont {Winn}, \citenamefont {Garlea}, \citenamefont {Abernathy},
  \citenamefont {Gao}, \citenamefont {Augustin}, \citenamefont {Santos},\ and\
  \citenamefont {Dai}}]{Chen2021}%
  \BibitemOpen
  \bibfield  {author} {\bibinfo {author} {\bibfnamefont {L.}~\bibnamefont
  {Chen}}, \bibinfo {author} {\bibfnamefont {J.-H.}\ \bibnamefont {Chung}},
  \bibinfo {author} {\bibfnamefont {M.~B.}\ \bibnamefont {Stone}}, \bibinfo
  {author} {\bibfnamefont {A.~I.}\ \bibnamefont {Kolesnikov}}, \bibinfo
  {author} {\bibfnamefont {B.}~\bibnamefont {Winn}}, \bibinfo {author}
  {\bibfnamefont {V.~O.}\ \bibnamefont {Garlea}}, \bibinfo {author}
  {\bibfnamefont {D.~L.}\ \bibnamefont {Abernathy}}, \bibinfo {author}
  {\bibfnamefont {B.}~\bibnamefont {Gao}}, \bibinfo {author} {\bibfnamefont
  {M.}~\bibnamefont {Augustin}}, \bibinfo {author} {\bibfnamefont {E.~J.~G.}\
  \bibnamefont {Santos}},\ and\ \bibinfo {author} {\bibfnamefont
  {P.}~\bibnamefont {Dai}},\ }\bibfield  {title} {\bibinfo {title} {{Magnetic
  Field Effect on Topological Spin Excitations in ${\mathrm{CrI}}_{3}$}},\
  }\href {https://doi.org/10.1103/PhysRevX.11.031047} {\bibfield  {journal}
  {\bibinfo  {journal} {Phys. Rev. X}\ }\textbf {\bibinfo {volume} {11}},\
  \bibinfo {pages} {031047} (\bibinfo {year} {2021})}\BibitemShut {NoStop}%
\bibitem [{\citenamefont {Zhu}\ \emph {et~al.}(2021)\citenamefont {Zhu},
  \citenamefont {Zhang}, \citenamefont {Wang}, \citenamefont {dos Santos},
  \citenamefont {Song}, \citenamefont {Mueller}, \citenamefont {Schmalzl},
  \citenamefont {Schmidt}, \citenamefont {Ivanov}, \citenamefont {Park},
  \citenamefont {Xu}, \citenamefont {Ma}, \citenamefont {Lounis}, \citenamefont
  {Bl\"{u}gel}, \citenamefont {Mokrousov}, \citenamefont {Su},\ and\
  \citenamefont {Br\"{u}ckel}}]{Zhu2021}%
  \BibitemOpen
  \bibfield  {author} {\bibinfo {author} {\bibfnamefont {F.}~\bibnamefont
  {Zhu}}, \bibinfo {author} {\bibfnamefont {L.}~\bibnamefont {Zhang}}, \bibinfo
  {author} {\bibfnamefont {X.}~\bibnamefont {Wang}}, \bibinfo {author}
  {\bibfnamefont {F.~J.}\ \bibnamefont {dos Santos}}, \bibinfo {author}
  {\bibfnamefont {J.}~\bibnamefont {Song}}, \bibinfo {author} {\bibfnamefont
  {T.}~\bibnamefont {Mueller}}, \bibinfo {author} {\bibfnamefont
  {K.}~\bibnamefont {Schmalzl}}, \bibinfo {author} {\bibfnamefont {W.~F.}\
  \bibnamefont {Schmidt}}, \bibinfo {author} {\bibfnamefont {A.}~\bibnamefont
  {Ivanov}}, \bibinfo {author} {\bibfnamefont {J.~T.}\ \bibnamefont {Park}},
  \bibinfo {author} {\bibfnamefont {J.}~\bibnamefont {Xu}}, \bibinfo {author}
  {\bibfnamefont {J.}~\bibnamefont {Ma}}, \bibinfo {author} {\bibfnamefont
  {S.}~\bibnamefont {Lounis}}, \bibinfo {author} {\bibfnamefont
  {S.}~\bibnamefont {Bl\"{u}gel}}, \bibinfo {author} {\bibfnamefont
  {Y.}~\bibnamefont {Mokrousov}}, \bibinfo {author} {\bibfnamefont
  {Y.}~\bibnamefont {Su}},\ and\ \bibinfo {author} {\bibfnamefont
  {T.}~\bibnamefont {Br\"{u}ckel}},\ }\bibfield  {title} {\bibinfo {title}
  {{Topological magnon insulators in two-dimensional van der Waals ferromagnets
  CrSiTe$_{3}$ and CrGeTe$_{3}$: Toward intrinsic gap-tunability}},\ }\href
  {https://doi.org/10.1126/sciadv.abi7532} {\bibfield  {journal} {\bibinfo
  {journal} {Science Advances}\ }\textbf {\bibinfo {volume} {7}},\ \bibinfo
  {pages} {eabi7532} (\bibinfo {year} {2021})}\BibitemShut {NoStop}%
\bibitem [{\citenamefont {Dzyaloshinsky}(1958)}]{DZYALOSHINSKY1958}%
  \BibitemOpen
  \bibfield  {author} {\bibinfo {author} {\bibfnamefont {I.}~\bibnamefont
  {Dzyaloshinsky}},\ }\bibfield  {title} {\bibinfo {title} {A thermodynamic
  theory of “weak” ferromagnetism of antiferromagnetics},\ }\href
  {https://doi.org/https://doi.org/10.1016/0022-3697(58)90076-3} {\bibfield
  {journal} {\bibinfo  {journal} {J. Phys. Chem. Solids}\ }\textbf {\bibinfo
  {volume} {4}},\ \bibinfo {pages} {241} (\bibinfo {year} {1958})}\BibitemShut
  {NoStop}%
\bibitem [{\citenamefont {Moriya}(1960)}]{Moriya1960}%
  \BibitemOpen
  \bibfield  {author} {\bibinfo {author} {\bibfnamefont {T.}~\bibnamefont
  {Moriya}},\ }\bibfield  {title} {\bibinfo {title} {Anisotropic superexchange
  interaction and weak ferromagnetism},\ }\href
  {https://doi.org/10.1103/PhysRev.120.91} {\bibfield  {journal} {\bibinfo
  {journal} {Phys. Rev.}\ }\textbf {\bibinfo {volume} {120}},\ \bibinfo {pages}
  {91} (\bibinfo {year} {1960})}\BibitemShut {NoStop}%
\bibitem [{\citenamefont {Wang}\ \emph {et~al.}(2017)\citenamefont {Wang},
  \citenamefont {Su},\ and\ \citenamefont {Wang}}]{Wang2017}%
  \BibitemOpen
  \bibfield  {author} {\bibinfo {author} {\bibfnamefont {X.~S.}\ \bibnamefont
  {Wang}}, \bibinfo {author} {\bibfnamefont {Y.}~\bibnamefont {Su}},\ and\
  \bibinfo {author} {\bibfnamefont {X.~R.}\ \bibnamefont {Wang}},\ }\bibfield
  {title} {\bibinfo {title} {Topologically protected unidirectional edge spin
  waves and beam splitter},\ }\href
  {https://doi.org/10.1103/PhysRevB.95.014435} {\bibfield  {journal} {\bibinfo
  {journal} {Phys. Rev. B}\ }\textbf {\bibinfo {volume} {95}},\ \bibinfo
  {pages} {014435} (\bibinfo {year} {2017})}\BibitemShut {NoStop}%
\bibitem [{\citenamefont {McClarty}\ \emph {et~al.}(2018)\citenamefont
  {McClarty}, \citenamefont {Dong}, \citenamefont {Gohlke}, \citenamefont
  {Rau}, \citenamefont {Pollmann}, \citenamefont {Moessner},\ and\
  \citenamefont {Penc}}]{McClarty2018}%
  \BibitemOpen
  \bibfield  {author} {\bibinfo {author} {\bibfnamefont {P.~A.}\ \bibnamefont
  {McClarty}}, \bibinfo {author} {\bibfnamefont {X.-Y.}\ \bibnamefont {Dong}},
  \bibinfo {author} {\bibfnamefont {M.}~\bibnamefont {Gohlke}}, \bibinfo
  {author} {\bibfnamefont {J.~G.}\ \bibnamefont {Rau}}, \bibinfo {author}
  {\bibfnamefont {F.}~\bibnamefont {Pollmann}}, \bibinfo {author}
  {\bibfnamefont {R.}~\bibnamefont {Moessner}},\ and\ \bibinfo {author}
  {\bibfnamefont {K.}~\bibnamefont {Penc}},\ }\bibfield  {title} {\bibinfo
  {title} {{Topological magnons in Kitaev magnets at high fields}},\ }\href
  {https://doi.org/10.1103/PhysRevB.98.060404} {\bibfield  {journal} {\bibinfo
  {journal} {Phys. Rev. B}\ }\textbf {\bibinfo {volume} {98}},\ \bibinfo
  {pages} {060404} (\bibinfo {year} {2018})}\BibitemShut {NoStop}%
\bibitem [{\citenamefont {Lee}\ \emph {et~al.}(2020)\citenamefont {Lee},
  \citenamefont {Utermohlen}, \citenamefont {Weber}, \citenamefont {Hwang},
  \citenamefont {Zhang}, \citenamefont {van Tol}, \citenamefont {Goldberger},
  \citenamefont {Trivedi},\ and\ \citenamefont {Hammel}}]{Lee2020}%
  \BibitemOpen
  \bibfield  {author} {\bibinfo {author} {\bibfnamefont {I.}~\bibnamefont
  {Lee}}, \bibinfo {author} {\bibfnamefont {F.~G.}\ \bibnamefont {Utermohlen}},
  \bibinfo {author} {\bibfnamefont {D.}~\bibnamefont {Weber}}, \bibinfo
  {author} {\bibfnamefont {K.}~\bibnamefont {Hwang}}, \bibinfo {author}
  {\bibfnamefont {C.}~\bibnamefont {Zhang}}, \bibinfo {author} {\bibfnamefont
  {J.}~\bibnamefont {van Tol}}, \bibinfo {author} {\bibfnamefont {J.~E.}\
  \bibnamefont {Goldberger}}, \bibinfo {author} {\bibfnamefont
  {N.}~\bibnamefont {Trivedi}},\ and\ \bibinfo {author} {\bibfnamefont {P.~C.}\
  \bibnamefont {Hammel}},\ }\bibfield  {title} {\bibinfo {title} {{Fundamental
  Spin Interactions Underlying the Magnetic Anisotropy in the Kitaev
  Ferromagnet ${\mathrm{CrI}}_{3}$}},\ }\href
  {https://doi.org/10.1103/PhysRevLett.124.017201} {\bibfield  {journal}
  {\bibinfo  {journal} {Phys. Rev. Lett.}\ }\textbf {\bibinfo {volume} {124}},\
  \bibinfo {pages} {017201} (\bibinfo {year} {2020})}\BibitemShut {NoStop}%
\bibitem [{\citenamefont {Jaeschke-Ubiergo}\ \emph {et~al.}(2021)\citenamefont
  {Jaeschke-Ubiergo}, \citenamefont {Su\'arez~Morell},\ and\ \citenamefont
  {Nunez}}]{Jaeschke-Ubiergo2021}%
  \BibitemOpen
  \bibfield  {author} {\bibinfo {author} {\bibfnamefont {R.}~\bibnamefont
  {Jaeschke-Ubiergo}}, \bibinfo {author} {\bibfnamefont {E.}~\bibnamefont
  {Su\'arez~Morell}},\ and\ \bibinfo {author} {\bibfnamefont {A.~S.}\
  \bibnamefont {Nunez}},\ }\bibfield  {title} {\bibinfo {title} {{Theory of
  magnetism in the van der Waals magnet ${\mathrm{CrI}}_{3}$}},\ }\href
  {https://doi.org/10.1103/PhysRevB.103.174410} {\bibfield  {journal} {\bibinfo
   {journal} {Phys. Rev. B}\ }\textbf {\bibinfo {volume} {103}},\ \bibinfo
  {pages} {174410} (\bibinfo {year} {2021})}\BibitemShut {NoStop}%
\bibitem [{\citenamefont {Brehm}\ \emph {et~al.}(2024)\citenamefont {Brehm},
  \citenamefont {Sobieszczyk}, \citenamefont {Kl\o{}getvedt}, \citenamefont
  {Evans}, \citenamefont {Santos},\ and\ \citenamefont
  {Qaiumzadeh}}]{Brehm2024}%
  \BibitemOpen
  \bibfield  {author} {\bibinfo {author} {\bibfnamefont {V.}~\bibnamefont
  {Brehm}}, \bibinfo {author} {\bibfnamefont {P.}~\bibnamefont {Sobieszczyk}},
  \bibinfo {author} {\bibfnamefont {J.~N.}\ \bibnamefont {Kl\o{}getvedt}},
  \bibinfo {author} {\bibfnamefont {R.~F.~L.}\ \bibnamefont {Evans}}, \bibinfo
  {author} {\bibfnamefont {E.~J.~G.}\ \bibnamefont {Santos}},\ and\ \bibinfo
  {author} {\bibfnamefont {A.}~\bibnamefont {Qaiumzadeh}},\ }\bibfield  {title}
  {\bibinfo {title} {{Topological magnon gap engineering in van der Waals
  ${\mathrm{CrI}}_{3}$ ferromagnets}},\ }\href
  {https://doi.org/10.1103/PhysRevB.109.174425} {\bibfield  {journal} {\bibinfo
   {journal} {Phys. Rev. B}\ }\textbf {\bibinfo {volume} {109}},\ \bibinfo
  {pages} {174425} (\bibinfo {year} {2024})}\BibitemShut {NoStop}%
\bibitem [{sup()}]{supp}%
  \BibitemOpen
  \href@noop {} {}\bibinfo {note} {See Supplemental Material at [URL] for
  considerations about angular-momentum conservation, the derivation of linear
  spin-wave theory and the Berry curvature in the inertial regime, a discussion
  of the parameter dependence of the size of the topological gap and of the
  topological phase transition in the honeycomb lattice, and the formation of a
  magnonic Chern insulating state in a single-sublattice ferromagnet. It also
  contains Refs.~\cite{Pommier1990,Allenspach1990} and the Supplemental
  Movies.}\BibitemShut {Stop}%
\bibitem [{\citenamefont {Scholtz}\ \emph {et~al.}(1992)\citenamefont
  {Scholtz}, \citenamefont {Geyer},\ and\ \citenamefont {Hahne}}]{Scholtz1992}%
  \BibitemOpen
  \bibfield  {author} {\bibinfo {author} {\bibfnamefont {F.}~\bibnamefont
  {Scholtz}}, \bibinfo {author} {\bibfnamefont {H.}~\bibnamefont {Geyer}},\
  and\ \bibinfo {author} {\bibfnamefont {F.}~\bibnamefont {Hahne}},\ }\bibfield
   {title} {\bibinfo {title} {{Quasi-Hermitian operators in quantum mechanics
  and the variational principle}},\ }\href
  {https://doi.org/10.1016/0003-4916(92)90284-s} {\bibfield  {journal}
  {\bibinfo  {journal} {Annals of Physics}\ }\textbf {\bibinfo {volume}
  {213}},\ \bibinfo {pages} {74–101} (\bibinfo {year} {1992})}\BibitemShut
  {NoStop}%
\bibitem [{\citenamefont {Mondal}\ \emph {et~al.}(2021)\citenamefont {Mondal},
  \citenamefont {Gro\ss{}enbach}, \citenamefont {R\'ozsa},\ and\ \citenamefont
  {Nowak}}]{Mondal_Ritwik_2021}%
  \BibitemOpen
  \bibfield  {author} {\bibinfo {author} {\bibfnamefont {R.}~\bibnamefont
  {Mondal}}, \bibinfo {author} {\bibfnamefont {S.}~\bibnamefont
  {Gro\ss{}enbach}}, \bibinfo {author} {\bibfnamefont {L.}~\bibnamefont
  {R\'ozsa}},\ and\ \bibinfo {author} {\bibfnamefont {U.}~\bibnamefont
  {Nowak}},\ }\bibfield  {title} {\bibinfo {title} {Nutation in
  antiferromagnetic resonance},\ }\href
  {https://doi.org/10.1103/PhysRevB.103.104404} {\bibfield  {journal} {\bibinfo
   {journal} {Phys. Rev. B}\ }\textbf {\bibinfo {volume} {103}},\ \bibinfo
  {pages} {104404} (\bibinfo {year} {2021})}\BibitemShut {NoStop}%
\bibitem [{\citenamefont {Soriano}\ \emph {et~al.}(2020)\citenamefont
  {Soriano}, \citenamefont {Katsnelson},\ and\ \citenamefont
  {Fernández-Rossier}}]{Soriano2020}%
  \BibitemOpen
  \bibfield  {author} {\bibinfo {author} {\bibfnamefont {D.}~\bibnamefont
  {Soriano}}, \bibinfo {author} {\bibfnamefont {M.~I.}\ \bibnamefont
  {Katsnelson}},\ and\ \bibinfo {author} {\bibfnamefont {J.}~\bibnamefont
  {Fernández-Rossier}},\ }\bibfield  {title} {\bibinfo {title} {Magnetic
  two-dimensional chromium trihalides: A theoretical perspective},\ }\href
  {https://doi.org/10.1021/acs.nanolett.0c02381} {\bibfield  {journal}
  {\bibinfo  {journal} {Nano Letters}\ }\textbf {\bibinfo {volume} {20}},\
  \bibinfo {pages} {6225–6234} (\bibinfo {year} {2020})}\BibitemShut
  {NoStop}%
\bibitem [{\citenamefont {Pommier}\ \emph {et~al.}(1990)\citenamefont
  {Pommier}, \citenamefont {Meyer}, \citenamefont {P\'enissard}, \citenamefont
  {Ferr\'e}, \citenamefont {Bruno},\ and\ \citenamefont
  {Renard}}]{Pommier1990}%
  \BibitemOpen
  \bibfield  {author} {\bibinfo {author} {\bibfnamefont {J.}~\bibnamefont
  {Pommier}}, \bibinfo {author} {\bibfnamefont {P.}~\bibnamefont {Meyer}},
  \bibinfo {author} {\bibfnamefont {G.}~\bibnamefont {P\'enissard}}, \bibinfo
  {author} {\bibfnamefont {J.}~\bibnamefont {Ferr\'e}}, \bibinfo {author}
  {\bibfnamefont {P.}~\bibnamefont {Bruno}},\ and\ \bibinfo {author}
  {\bibfnamefont {D.}~\bibnamefont {Renard}},\ }\bibfield  {title} {\bibinfo
  {title} {Magnetization reversal in ultrathin ferromagnetic films with
  perpendicular anistropy: Domain observations},\ }\href
  {https://doi.org/10.1103/PhysRevLett.65.2054} {\bibfield  {journal} {\bibinfo
   {journal} {Phys. Rev. Lett.}\ }\textbf {\bibinfo {volume} {65}},\ \bibinfo
  {pages} {2054} (\bibinfo {year} {1990})}\BibitemShut {NoStop}%
\bibitem [{\citenamefont {Allenspach}\ \emph {et~al.}(1990)\citenamefont
  {Allenspach}, \citenamefont {Stampanoni},\ and\ \citenamefont
  {Bischof}}]{Allenspach1990}%
  \BibitemOpen
  \bibfield  {author} {\bibinfo {author} {\bibfnamefont {R.}~\bibnamefont
  {Allenspach}}, \bibinfo {author} {\bibfnamefont {M.}~\bibnamefont
  {Stampanoni}},\ and\ \bibinfo {author} {\bibfnamefont {A.}~\bibnamefont
  {Bischof}},\ }\bibfield  {title} {\bibinfo {title} {{Magnetic domains in thin
  epitaxial Co/Au(111) films}},\ }\href
  {https://doi.org/10.1103/PhysRevLett.65.3344} {\bibfield  {journal} {\bibinfo
   {journal} {Phys. Rev. Lett.}\ }\textbf {\bibinfo {volume} {65}},\ \bibinfo
  {pages} {3344} (\bibinfo {year} {1990})}\BibitemShut {NoStop}%
\end{thebibliography}

\begin{thebibliography}{8}%
\setcounter{NAT@ctr}{0}
\renewcommand{\bibnumfmt}[1]{[S#1]}
\makeatletter
\providecommand \@ifxundefined [1]{%
 \@ifx{#1\undefined}
}%
\providecommand \@ifnum [1]{%
 \ifnum #1\expandafter \@firstoftwo
 \else \expandafter \@secondoftwo
 \fi
}%
\providecommand \@ifx [1]{%
 \ifx #1\expandafter \@firstoftwo
 \else \expandafter \@secondoftwo
 \fi
}%
\providecommand \natexlab [1]{#1}%
\providecommand \enquote  [1]{``#1''}%
\providecommand \bibnamefont  [1]{#1}%
\providecommand \bibfnamefont [1]{#1}%
\providecommand \citenamefont [1]{#1}%
\providecommand \href@noop [0]{\@secondoftwo}%
\providecommand \href [0]{\begingroup \@sanitize@url \@href}%
\providecommand \@href[1]{\@@startlink{#1}\@@href}%
\providecommand \@@href[1]{\endgroup#1\@@endlink}%
\providecommand \@sanitize@url [0]{\catcode `\\12\catcode `\$12\catcode
  `\&12\catcode `\#12\catcode `\^12\catcode `\_12\catcode `\%12\relax}%
\providecommand \@@startlink[1]{}%
\providecommand \@@endlink[0]{}%
\providecommand \url  [0]{\begingroup\@sanitize@url \@url }%
\providecommand \@url [1]{\endgroup\@href {#1}{\urlprefix }}%
\providecommand \urlprefix  [0]{URL }%
\providecommand \Eprint [0]{\href }%
\providecommand \doibase [0]{http://dx.doi.org/}%
\providecommand \selectlanguage [0]{\@gobble}%
\providecommand \bibinfo  [0]{\@secondoftwo}%
\providecommand \bibfield  [0]{\@secondoftwo}%
\providecommand \translation [1]{[#1]}%
\providecommand \BibitemOpen [0]{}%
\providecommand \bibitemStop [0]{}%
\providecommand \bibitemNoStop [0]{.\EOS\space}%
\providecommand \EOS [0]{\spacefactor3000\relax}%
\providecommand \BibitemShut  [1]{\csname bibitem#1\endcsname}%
\let\auto@bib@innerbib\@empty
\bibitem [{\citenamefont {Ciornei}\ \emph {et~al.}(2011)\citenamefont
  {Ciornei}, \citenamefont {Rub\'{\i}},\ and\ \citenamefont
  {Wegrowe}}]{Ciornei2011_supp}%
  \BibitemOpen
  \bibfield  {author} {\bibinfo {author} {\bibfnamefont {M.-C.}\ \bibnamefont
  {Ciornei}}, \bibinfo {author} {\bibfnamefont {J.~M.}\ \bibnamefont
  {Rub\'{\i}}}, \ and\ \bibinfo {author} {\bibfnamefont {J.-E.}\ \bibnamefont
  {Wegrowe}},\ }\bibfield  {title} {\enquote {\bibinfo {title} {Magnetization
  dynamics in the inertial regime: Nutation predicted at short time scales},}\
  }\href {\doibase 10.1103/PhysRevB.83.020410} {\bibfield  {journal} {\bibinfo
  {journal} {Phys. Rev. B}\ }\textbf {\bibinfo {volume} {83}},\ \bibinfo
  {pages} {020410} (\bibinfo {year} {2011})}\BibitemShut {NoStop}%
\bibitem [{\citenamefont {Scholtz}\ \emph {et~al.}(1992)\citenamefont
  {Scholtz}, \citenamefont {Geyer},\ and\ \citenamefont {Hahne}}]{Scholtz1992_supp}%
  \BibitemOpen
  \bibfield  {author} {\bibinfo {author} {\bibfnamefont {F.G.}\ \bibnamefont
  {Scholtz}}, \bibinfo {author} {\bibfnamefont {H.B.}\ \bibnamefont {Geyer}}, \
  and\ \bibinfo {author} {\bibfnamefont {F.J.W.}\ \bibnamefont {Hahne}},\
  }\bibfield  {title} {\enquote {\bibinfo {title} {{Quasi-Hermitian operators
  in quantum mechanics and the variational principle}},}\ }\href {\doibase
  10.1016/0003-4916(92)90284-s} {\bibfield  {journal} {\bibinfo  {journal}
  {Annals of Physics}\ }\textbf {\bibinfo {volume} {213}},\ \bibinfo {pages}
  {74–101} (\bibinfo {year} {1992})}\BibitemShut {NoStop}%
\bibitem [{\citenamefont {Berry}(1984)}]{Berry1984_supp}%
  \BibitemOpen
  \bibfield  {author} {\bibinfo {author} {\bibfnamefont {Michael~Victor}\
  \bibnamefont {Berry}},\ }\bibfield  {title} {\enquote {\bibinfo {title}
  {Quantal phase factors accompanying adiabatic changes},}\ }\href {\doibase
  10.1098/rspa.1984.0023} {\bibfield  {journal} {\bibinfo  {journal}
  {Proceedings of the Royal Society of London. A. Mathematical and Physical
  Sciences}\ }\textbf {\bibinfo {volume} {392}},\ \bibinfo {pages} {45–57}
  (\bibinfo {year} {1984})}\BibitemShut {NoStop}%
\bibitem [{\citenamefont {Zhang}\ \emph {et~al.}(2013)\citenamefont {Zhang},
  \citenamefont {Ren}, \citenamefont {Wang},\ and\ \citenamefont
  {Li}}]{Zhang2013_supp}%
  \BibitemOpen
  \bibfield  {author} {\bibinfo {author} {\bibfnamefont {Lifa}\ \bibnamefont
  {Zhang}}, \bibinfo {author} {\bibfnamefont {Jie}\ \bibnamefont {Ren}},
  \bibinfo {author} {\bibfnamefont {Jian-Sheng}\ \bibnamefont {Wang}}, \ and\
  \bibinfo {author} {\bibfnamefont {Baowen}\ \bibnamefont {Li}},\ }\bibfield
  {title} {\enquote {\bibinfo {title} {Topological magnon insulator in
  insulating ferromagnet},}\ }\href {\doibase 10.1103/PhysRevB.87.144101}
  {\bibfield  {journal} {\bibinfo  {journal} {Phys. Rev. B}\ }\textbf {\bibinfo
  {volume} {87}},\ \bibinfo {pages} {144101} (\bibinfo {year}
  {2013})}\BibitemShut {NoStop}%
\bibitem [{\citenamefont {Shindou}\ \emph {et~al.}(2013)\citenamefont
  {Shindou}, \citenamefont {Matsumoto}, \citenamefont {Murakami},\ and\
  \citenamefont {Ohe}}]{Shindou2013_supp}%
  \BibitemOpen
  \bibfield  {author} {\bibinfo {author} {\bibfnamefont {Ryuichi}\ \bibnamefont
  {Shindou}}, \bibinfo {author} {\bibfnamefont {Ryo}\ \bibnamefont
  {Matsumoto}}, \bibinfo {author} {\bibfnamefont {Shuichi}\ \bibnamefont
  {Murakami}}, \ and\ \bibinfo {author} {\bibfnamefont {Jun-ichiro}\
  \bibnamefont {Ohe}},\ }\bibfield  {title} {\enquote {\bibinfo {title}
  {Topological chiral magnonic edge mode in a magnonic crystal},}\ }\href
  {\doibase 10.1103/PhysRevB.87.174427} {\bibfield  {journal} {\bibinfo
  {journal} {Phys. Rev. B}\ }\textbf {\bibinfo {volume} {87}},\ \bibinfo
  {pages} {174427} (\bibinfo {year} {2013})}\BibitemShut {NoStop}%
\bibitem [{\citenamefont {McClarty}(2022)}]{mcclarty2022_supp}%
  \BibitemOpen
  \bibfield  {author} {\bibinfo {author} {\bibfnamefont {Paul~A.}\ \bibnamefont
  {McClarty}},\ }\bibfield  {title} {\enquote {\bibinfo {title} {Topological
  magnons: A review},}\ }\href {\doibase
  https://doi.org/10.1146/annurev-conmatphys-031620-104715} {\bibfield
  {journal} {\bibinfo  {journal} {Annual Review of Condensed Matter Physics}\
  }\textbf {\bibinfo {volume} {13}},\ \bibinfo {pages} {171--190} (\bibinfo
  {year} {2022})}\BibitemShut {NoStop}%
\bibitem [{\citenamefont {Pommier}\ \emph {et~al.}(1990)\citenamefont
  {Pommier}, \citenamefont {Meyer}, \citenamefont {P\'enissard}, \citenamefont
  {Ferr\'e}, \citenamefont {Bruno},\ and\ \citenamefont
  {Renard}}]{Pommier1990_supp}%
  \BibitemOpen
  \bibfield  {author} {\bibinfo {author} {\bibfnamefont {J.}~\bibnamefont
  {Pommier}}, \bibinfo {author} {\bibfnamefont {P.}~\bibnamefont {Meyer}},
  \bibinfo {author} {\bibfnamefont {G.}~\bibnamefont {P\'enissard}}, \bibinfo
  {author} {\bibfnamefont {J.}~\bibnamefont {Ferr\'e}}, \bibinfo {author}
  {\bibfnamefont {P.}~\bibnamefont {Bruno}}, \ and\ \bibinfo {author}
  {\bibfnamefont {D.}~\bibnamefont {Renard}},\ }\bibfield  {title} {\enquote
  {\bibinfo {title} {Magnetization reversal in ultrathin ferromagnetic films
  with perpendicular anistropy: Domain observations},}\ }\href {\doibase
  10.1103/PhysRevLett.65.2054} {\bibfield  {journal} {\bibinfo  {journal}
  {Phys. Rev. Lett.}\ }\textbf {\bibinfo {volume} {65}},\ \bibinfo {pages}
  {2054--2057} (\bibinfo {year} {1990})}\BibitemShut {NoStop}%
\bibitem [{\citenamefont {Allenspach}\ \emph {et~al.}(1990)\citenamefont
  {Allenspach}, \citenamefont {Stampanoni},\ and\ \citenamefont
  {Bischof}}]{Allenspach1990_supp}%
  \BibitemOpen
  \bibfield  {author} {\bibinfo {author} {\bibfnamefont {R.}~\bibnamefont
  {Allenspach}}, \bibinfo {author} {\bibfnamefont {M.}~\bibnamefont
  {Stampanoni}}, \ and\ \bibinfo {author} {\bibfnamefont {A.}~\bibnamefont
  {Bischof}},\ }\bibfield  {title} {\enquote {\bibinfo {title} {{Magnetic
  domains in thin epitaxial Co/Au(111) films}},}\ }\href {\doibase
  10.1103/PhysRevLett.65.3344} {\bibfield  {journal} {\bibinfo  {journal}
  {Phys. Rev. Lett.}\ }\textbf {\bibinfo {volume} {65}},\ \bibinfo {pages}
  {3344--3347} (\bibinfo {year} {1990})}\BibitemShut {NoStop}%
\end{thebibliography}
\end{document}